%% file: main.tex
\documentclass[lettersize,journal,hidelinks]{IEEEtran}

\usepackage[table, usenames, dvipsnames]{xcolor}

\usepackage[commentmarkup=uwave, final]{changes}
\definechangesauthor[name=Marc, color=orange]{ml}
\definechangesauthor[name=Andreas, color=red]{ah}
\definechangesauthor[name=Chris, color=violet]{cs}
\definechangesauthor[name=Konstantin, color=purple]{kk}
\usepackage{amsmath,amsfonts}
\usepackage{algorithmic}
\usepackage{algorithm}
\usepackage{array}
\usepackage{textcomp}
\usepackage{stfloats}
\usepackage{url}
\usepackage{verbatim}
\usepackage{graphicx}
\usepackage{cite}

\usepackage{pgf}
\usepackage{svg}
\usepackage{microtype}
\usepackage{times}

\usepackage{multirow}
\usepackage{booktabs}
\usepackage{tabu}
\usepackage{hyperref}
\usepackage{cleveref}
\usepackage{tabularx}
\usepackage{mathptmx}
\usepackage{siunitx}
\usepackage{subfig}
\usepackage{todonotes}
\newcommand{\acc}[2]{$Acc_{\text{enr}: #1}^{\text{use}: #2}$}

\begin{document}
\title{Versatile User Identification in Extended Reality using Pretrained Similarity-Learning}

\author{Christian Rack, Konstantin Kobs, Tamara Fernando, Andreas Hotho, Marc Erich Latoschik}

\maketitle


\begin{abstract}
Various machine learning approaches have proven to be useful for user verification and identification based on motion data in eXtended Reality (XR). However, their real-world application still faces significant challenges concerning versatility, i.e., in terms of extensibility and generalization capability.
This article presents a solution that is both extensible to new users without expensive retraining, and that generalizes well across different sessions, devices, and user tasks.
To this end, we developed a similarity-learning model and pretrained it on the ``Who~Is~Alyx?" dataset.
This dataset features a wide array of tasks and hence motions from users playing the VR game ``Half-Life: Alyx".
In contrast to previous works, we used a dedicated set of users for model validation and final evaluation.
Furthermore, we extended this evaluation using an independent dataset that features completely different users, tasks, and three different XR devices.
In comparison with a traditional classification-learning baseline, our model shows superior performance, especially in scenarios with limited enrollment data.
The pretraining process allows immediate deployment in a diverse range of XR applications while maintaining high versatility.
Looking ahead, our approach paves the way for easy integration of pretrained motion-based identification models in production XR systems.
\end{abstract}


\input{sections/01_introduction}
\input{sections/02_related_work}
\input{sections/03_methods}
\input{sections/04_results}
\input{sections/05_discussion}

\bibliographystyle{ieeetr}
\bibliography{references,chris_zotero_export}
\end{document}

%% file: sections/01_introduction.tex

\section{Introduction}

\begin{figure*}[b]
    \centering
    \includegraphics[width=1\linewidth]{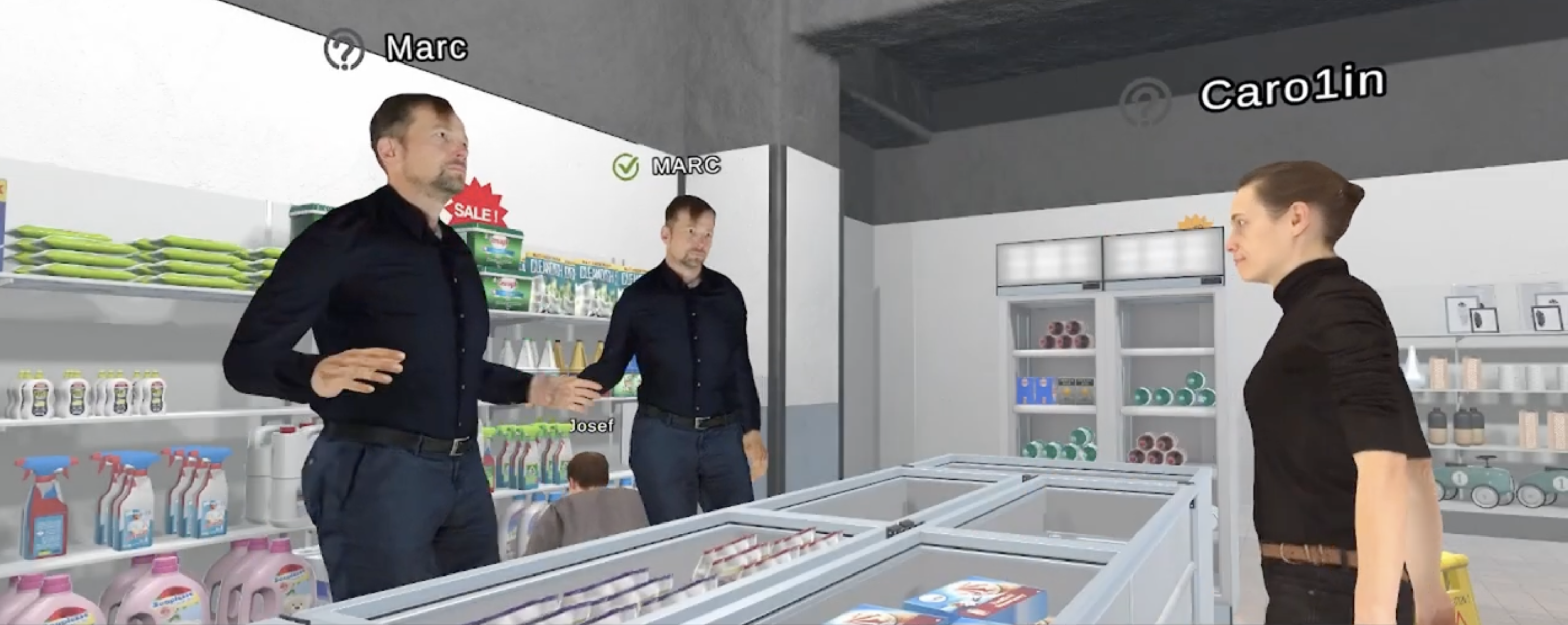}
    \caption{
     Who is the real Marc? This Social VR scene taken from Lin et al. \cite{Lin2023} demonstrates how a versatile identification model can be used to reveal impostors and, hence, foster trust in SocialVR applications: two users claim to be `Marc', but, as the identification system indicates, only one is genuine.}
    \label{fig:teaser}
\end{figure*}

Virtual, Augmented, and Mixed Reality (VR, AR, MR: eXtended Reality, or XR for short) provide many novel and interesting ways to interact with others in a future Metaverse.
Arguably the most defining feature of such a Metaverse is based on the idea of a Social XR:
users engage with each other through avatars, their digital alter egos in the virtual realm.
Avatar embodiment promises to make the whole range of verbal and non-verbal social signals accessible in virtual encounters, from the full range of gestures to even sometimes just subtleties of facial expressions ~\cite{smith2018communication, Latoschik2019,latoschik2017effect}.

However, the embodiment by avatars in social virtual encounters brings with it considerable challenges in terms of privacy, identity, and trust~\cite{lin2022digital}.
How do we know with whom we really interact, i.e., who is controlling the avatars of our virtual peers?
This problem is significantly exacerbated by the advances in automatic 3D reconstruction of photorealistic avatars~\cite{achenbach2017generation,conf/vrst/WenningerABLB20, 2021:bartl:affordable}, elevating risks of impersonation, privacy breaches, and fraud~\cite{lin2022digital}, and the overall question of a right to one's own virtual body and how such a right can be secured.
Such questions about virtual doppelgangers are already discussed in the movie industry.
Metaverse technologies drastically aggravate this problem since they raise similar questions around user identity and authenticity for real-time interactive encounters.

Consequently, ensuring user authenticity is crucial for maintaining the integrity and trustworthiness of Social XR spaces.
This necessitates robust user identification and verification mechanisms that can operate unobtrusively within the immersive experience.
In this context, it is interesting to investigate the potential of users' motions.
Motion tracking is a central feature of XR setups.
Even consumer-grade XR systems track the user's head position and orientation (to update the egocentric view) and at least one hand or controller (to approximate the position and orientation of a hand), since this is essential for creating an immersive experience and facilitating system interactions.
Recent work \cite{Kupin2019, Li2016, Liebers2020, Liebers2021, Liebers2022, Mathis2020, Miller2019, Miller2020, Miller2020a, Miller2021, Miller2022CombiningBiometrics, Miller2022TemporalBiometrics, Moore2021PersonalSessions, Munsell2012, Nguyen2018, Pfeuffer2019, Rogers2015, Rack2022ComparisonSequences} has shown that this tracking data contains identifying patterns of users.
Hence, it can be exploited for novel biometric motion-based identification and verification (i.e., authentication) solutions:
instead of having to deal with cumbersome keyboards for password-based methods or additional hardware (i.e., iris or fingerprint scanner), XR users can be recognized solely based on their individual motion patterns.
In addition, this recognition is not only applicable during onboarding (i.e., login).
Instead, it can be performed even unobtrusively and at any time during an XR session, enabling on-demand verification of our virtual peers.

For a motion-based identification and verification model to be broadly applicable in real-world Metaverse scenarios, its versatility becomes paramount.
Versatility encompasses several critical characteristics: the model's identification accuracy, its extensibility to new users, and its ability to generalize across various signal characteristics and sources of noise:
in general, we must prevent our models from overfitting to non user-specific patterns as potentially elicited by specific tasks or different devices.
We are primarily interested in the user-specific motion characteristics independent from task, device, or session.

The accuracy of motion-based identification models has been extensively explored, and some works have also delved into aspects of generalization.
However, previous works have focused predominantly on motion-based identification models that are specialized in a limited set of XR users, which raises the question about the practicality of these systems in broader real-world scenarios.
In such a context, the need for versatile models that can quickly adapt to new users without requiring deep machine learning expertise or significant computational resources becomes apparent.
Hence, in this paper we investigate the potential of a similarity-learning architecture that promises to be fully versatile, as once pretrained it can be used to identify new users without further retraining across different settings.

While there have been first steps to explore the feasibility of the similarity-learning approach \cite{Miller2022CombiningBiometrics}, the versatility challenges have not been comprehensively addressed so far.
Primarily, it remains unexplored whether these models retain their effectiveness when applied to a completely new set of users, i.e., users whom the model has not been trained on.
Additionally, it is unclear whether such models also work in situations where users are not performing just a very specific action, but where motions are very different and unpredictable.

Our work addresses these gaps by developing a pretrained similarity-learning model on the ``Who is Alyx?" dataset which provides tracking data of VR users performing a wide array of tasks and hence resulting motion patterns. 
For this, we use separate sets of users for the pretraining and testing phases, as only this allows us to examine the pretrained model's potential to generalize to previously unseen individuals.
Our analysis includes a comparison with a state-of-the-art classification-learning model as baseline, trained and evaluated on the same test users.
Following a thorough hyperparameter search for both models, the results show that the baseline model requires more data to achieve comparable performance to our pretrained similarity-learning model on the test set.
In addition, we further extend our model's validation using an independent alternative dataset from Miller R. et al~\cite{Miller2022CombiningBiometrics}.
This dataset features completely different users, tasks, and three different types of XR devices.
With this setup, we demonstrate that our similarity-learning model, once pretrained on the ``Who~Is~Alyx?'' dataset, also generalizes well across users, tasks, and devices.

Altogether, this paper demonstrates that the similarity-learning approach remains effective, even if pretrained on an entirely different set of users and used in a scenario with arbitrary and unpredictable user motions.
This contributes a novel perspective on the versatility of similarity-learning models for motion-based user identification in XR and holds the potential to bridge the divide between theoretical discussion and practical application in the Metaverse.

%% file: sections/02_related_work.tex
\section{Related Work}
\definecolor{lightgray}{gray}{0.95}
\begin{table*}[]
\centering
\caption{Comparison of motion-based user identification models by their demonstrated versatility; the user task of the evaluated dataset and the capabilities of the model type determine the various aspects of versatility (accuracy, onboarding of new users and generalization capabilities across tasks, sessions and devices) that could be demonstrated by the individual works.}
\label{tab:previous-work}
\begin{tabular}{lll||ccccc}
\toprule
\textbf{Publication}                                        & \textbf{User Task}             & \textbf{Model Type} & \multicolumn{5}{c}{\textbf{Demonstrated Versatility} } \\
                                                            &                                &                     & \textbf{Accuracy}   & \multicolumn{3}{c}{\textbf{Generalization Across}} & \textbf{New Users}  \\
                                                            &                                &                     &                     & Sessions       & Devices & Tasks                   &                     \\[0.25cm]\midrule
\multicolumn{8}{l}{\textbf{Pretrainable Methods}} \\[0.1cm]
\rowcolor{lightgray}
\hspace{6.5mm}This work                                     & playing Half-Life: Alyx        & similarity-learn.   & +                   & +              & + 		& ++                      & +                   \\
\cite{Miller2021} Miller R et al. (2021)                    & throwing a ball                & similarity-learn.   & +                   & +              & + 		&                         & +                   \\
\cite{Kupin2019} Kupin et al. (2019)                        & throwing a ball                & feature-distance    & +                   & +              &   		&                         &                     \\
\cite{Li2016} Li et al. (2016)                              & nodding to music               & feature-distance    & +                   & +              &   		&                         &                     \\[0.25cm]\midrule\midrule
\multicolumn{8}{l}{\textbf{Non-Pretrainable Methods}} \\[0.1cm]                                                                                                                                                                             
\cite{RackWhoAlyxnew2023} Rack et al. (2023)                          & playing Half-Life: Alyx        & class.-learn.       & +                   & +              &         & ++                      &                     \\
\cite{moore_identifying_2023} Moore et al. (2023)           & construction assembly          & class.-learn.       & +                   & +              &         & +                       &                     \\
\cite{Liebers2023} Liebers et al. (2023)                    & playing Beat Saber             & class.-learn.       & +                   & +              &         & +                       &                     \\
\cite{NairUniqueIdentification502023} Nair et al. (2023)                          & playing Beat Saber             & class.-learn.       & +                   & +              & +       & +                       &                     \\
\cite{Liebers2022} Liebers et al. (2022)                    & interacting with 3D UI         & class.-learn.       & +                   & +              &         &                         &                     \\
\cite{Quintero2021} Quntero et al. (2021)                   & watching 360° videos           & class.-learn.       & +                   &                &         &                         &                                                                     \\
\cite{Moore2021} Moore et al. (2021)                        & VR robot troubleshooting       & class.-learn.       & +                   & +              &         & +                       &                     \\
\cite{Bhalla2021} Bhalla et al. (2021)                      & typing on AR keyboard          & class.-learn.       & +                   & +              &         &                         &                     \\
\cite{Liebers2021} Liebers et al. (2021)                    & bowling, archery               & class.-learn.       & +                   & +              &         &                         &                     \\
\cite{Mathis2020} Mathis et al. (2020)                      & interaction with 3D cube       & class.-learn.       & +                   &                &         &                         &                                                                     \\
\cite{Olade2020} Olade et al. (2020)                        & moving objects                 & class.-learn.       & +                   & +              &         &                         &                     \\
\cite{Miller2020} Miller R. et al. (2020)                   & ball throwing                  & class.-learn.       & +                   & +              & +       &                         &                     \\
\cite{Miller2020a} Miller M. et al. (2020)                  & watching 360° videos           & class.-learn.       & +                   &                &         &                         &                                                                     \\
\cite{Ajit2019} Ajit et al. (2019)                          & throwing a ball                & class.-learn.       & +                   & +              &         &                         &                     \\
\cite{Pfeuffer2019} Pfeuffer et al. (2019)                  & point, grab, walk, type        & class.-learn.       & +                   & +              &         &                         &                     \\
\cite{Shen2019} Shen et al. (2019)                          & walking a few steps            & class.-learn.       & +                   & +              &         &                         &                     \\
\cite{Mustafa2018} Mustafa et al. (2018)                    & walking                        & class.-learn.       & +                   & +              &         &                         &                     \\
\cite{Rogers2015} Rogers et al. (2015)                      & watching videos                & class.-learn.       & +                   &                &         &                         &                                                                     \\ \bottomrule
\end{tabular}        
\end{table*}

There is a sizeable body of related work investigating motions of XR users as input for biometric systems.
In this section we discuss the current landscape of previous works and highlight our contributions.

\subsection{Versatility of Motion-Based Identification Methods}

To contextualize the contributions of the presented research, we assess prior solutions based on their versatility, as summarized in \Cref{tab:previous-work}.
We characterize a versatile identification model by its \textit{accuracy}, its ability to \textit{generalize} across diverse settings, and its \textit{extensibility} to onboard new users efficiently.
Accuracy in models refers to their capability to recognize users with high reliability for their intended use-cases.
Given the challenges of conducting a direct and fair comparison of identification accuracy across different works without a uniform dataset, we assume that previous models achieve adequate accuracy for their respective use-cases and refrain from further distinction of identification performances.
Models that generalize well can be applied to a broader range of use-cases, for example, to cases where devices can be different, user recordings are from separate sessions or where user actions can be arbitrary and unpredictable.
\Cref{sec:signal-vs-noise} provides a detailed discussion of generalization and the three types we distinguish in the context of motion-based identification.

Moreover, an ideal model is not restricted to recognizing only users encountered during training.
Instead, it supports pretraining on extensive datasets, allowing for subsequent application to new users without the need for comprehensive retraining.
This pretrainability is a crucial attribute, as it determines the utility of the motion-based identification approach in real-world situations where updating model parameters for every new user is impractically resource-intensive.
We frame the current landscape of motion-based identification approaches between pretrainable and non-pretrainable methods in \Cref{sec:distance_vs_classification}.


\subsection{Generalization Capabilities of Identification Models}
\label{sec:signal-vs-noise}
Central qualities of a model are determined by its accuracy to isolate the relevant signal from irrelevant noise within the incoming stream of data.
The \textit{relevant signal} in our context refers to user-specific characteristics, which are influenced by the individual's physical attributes, musculoskeletal system, possible limitations, overall fitness, and movement history.
A model that generalizes well can detect this signal even within a high amount of \textit{irrelevant noise} and can thus be deployed to a broader range of use-cases.

For our work, we distinguish three different types of noise: task-, session- and device-induced noise, which will be discussed in the following.
The ability of a model to handle either type of noise depends not only on its architecture, but also on the dataset it has been trained and evaluated on.
For example, if a dataset does not include more than one session per user, it cannot be assessed if the model can recognize the user again on a different day.
\Cref{tab:previous-work} provides an overview over the evaluated types of noise for each previous work, and derives the overall versatility of the proposed solutions based on this.

Early research primarily focused on establishing the existence of a user-specific signal \cite{Rogers2015,Li2016}.
Subsequent works then shifted towards developing strategies for managing the different types of noise to enhance model versatility.
Several publications \cite{Mustafa2018,Kupin2019,Ajit2019,Miller2021,Liebers2021,NairUniqueIdentification502023,RackWhoAlyxnew2023} regard session-induced noise, which stems from conditions specific to the current session, such as the user's position, orientation, or system calibration.
If a model overfits to these session-specific conditions, it may fail to re-identify the same user under altered conditions.
To evaluate if this happens, datasets need to provide recordings from not only one, but at least two sessions from the same user.
Rack et al. \cite{Rack2022ComparisonSequences} have shown that appropriate data encoding during preprocessing is an efficient way to mitigate this session-specific noise.

More recently, Miller R. et al. \cite{Miller2021} have started to investigate device-induced noise.
Different hardware, software and calibrations can induce individual patterns, which are not user-specific, but inherent to the XR setup.
The authors showed that the similarity-learning approach can effectively handle this form of noise, a finding we reproduce in our analyses.

The final and possibly most significant dimension affecting the generalization capability is defined by the task users are performing.
This noise emerges from the movements a user may perform within a given task.
The ideal model would be task-independent, accurately recognizing the user's unique motion patterns regardless of the specific actions being performed, and not overfitting to specific movements which are necessary to actually perform specific tasks.

To achieve this, it is essential to select datasets with a high task diversity for training and evaluation, as it determines the model's exposure to diverse motion patterns and its consequent ability to isolate the user-specific signal from the task-induced noise.
\Cref{tab:previous-work} provides an overview of the individual user task of each work's dataset, which we categorize in the following into user tasks with low, moderate and high diversity.
This diversity determines the model's demonstrated ability to work across different user tasks.
The following elaborates on the nuances of these levels of task diversity and their impact on model performance.

In scenarios with low task diversity, motions are consistent and well-defined.
Consequently, the task-induced noise is minimal, potentially making it easier for the model to isolate and identify the user-specific signal.
Examples of such use-cases include the works by Miller M. et al.\cite{Miller2020a} (``watching 360° videos''), Miller R. et al.\cite{Miller2021} (``throwing a ball''), and others \cite{Ajit2019,Kupin2019,Liebers2021,Olade2020,Pfeuffer2019,Shen2019,Mustafa2018}, where users performed very specific tasks.
These scenarios allow identification models to specialize on individual characteristics within predictable trajectories.
For example, any given sequence taken from the dataset of Miller R. et al.\cite{Miller2021} will include a ball-throwing motion.
On the one hand, such scenarios are highly relevant for verification use-cases, where users can be asked to actively provide a specific motion to the system.
On the other hand, these models are likely to be very task-dependent, so their overall versatility is limited.

Scenarios with a moderate task-diversity include a wider range of possible motions and present a higher degree of task-induced noise, posing a moderately complex challenge for user identification.
Works in this category, such as those by Rack et al. \cite{Rack2022ComparisonSequences}, Nair et al.\cite{NairUniqueIdentification502023}, and Moore et al.\cite{Moore2021PersonalSessions}, explore open-ended interactions like users talking to each other, playing the rhythm game ``Beat Saber'', or engaging in VR e-learning environments.
Here, users are typically given tasks allowing greater freedom of movement, introducing a broader spectrum of motion into the input data.
For example, sequences taken from the dataset of Nair et al.\cite{NairUniqueIdentification502023} will include mostly up- and down-motions of the arms, recurringly left- and right swipes, and occasionally squatting.
In this instance the model does not know in advance which motion it encounters in any given input sequence, but the scope of possible motions is still relatively limited.
Hence, models trained on such datasets may generalize to some extent across tasks, but only promise to work well within the scope of the training set.

Scenarios with a high task diversity allow a high degree of freedom to users and exhibit a wide range of motions, significantly complicating the task of differentiating the user-specific signal from the task-induced noise.
Here, the identification task is notably challenging, as the model must find user-specific signals within a vast array of potential motions.
Our work focuses on this category, utilizing the `Who Is Alyx?' dataset, which includes diverse user actions within a VR game environment.
Here, input sequences might show motions coming from the user calmly navigating through the level, energetically trying to solve a puzzle, or frantically fighting enemies.
Identification models working well in such diverse scenarios promise to be task-independent ––– an important quality especially for continuous user recognition use-cases, where users can pursue different tasks at any given moment.

Altogether, we investigate the similarity-learning model's ability to handle all three types of noise.
By employing a multi-session dataset that encompasses a wide variety of user actions, we explore the similarity-model's ability to handle different tasks and sessions.
Additionally, we use the dataset from Miller R. et al. \cite{Miller2022TemporalBiometrics} in our analyses to demonstrate the model's ability to work across different devices.

\subsection{Pretrainable and Non-Pretrainable Methods}
\label{sec:distance_vs_classification}

Machine learning methodologies for user identification in XR can be broadly categorized into pretrainable and non-pretrainable approaches.
Pretrainable model types, such as feature-distance and similarity-learning, offer significant advantages in terms of ease of deployment and extensibility to new users, which is crucial for practical applications.
On the other hand, non-pretrainable methods, primarily classification-learning models, require retraining for each new user, presenting a substantial barrier to their widespread adoption.
\Cref{tab:previous-work} categorizes previous works with regards to their pretrainability and notes which model type was used.

Particularly earlier works utilize \textit{feature-distance} methods to leverage XR user motions for identification purposes~\cite{Li2016,Shen2019,Kupin2019}.
These models make predictions by simple comparison of motion trajectories using distance metrics directly in the feature space, hence, without the need for a training phase.
While this would usually not be referred to as machine learning in the traditional sense, we included them under \Cref{tab:previous-work} as `pretrainable' methods for their ability to instantly onboard new users.
A new user simply provides a few reference motions, which are then added to the reference set for future inferences.
However, feature-distance models only work well for very a specific motion, so they do not generalize across different user tasks, restricting their overall versatility.

\textit{Classification-learning} models, in contrast, are not readily pretrainable.
Recent works have demonstrated their enhanced generalization capability \cite{RackWhoAlyxnew2023,NairUniqueIdentification502023}.
However, as they discern distinctive patterns in user behaviors even in noisy environments, they are limited to the set of users they were trained on.
Incorporating new users requires recording of substantial amounts of new the users' motion data followed by  extensive model training, a process that demands time and resources.
This poses challenges for XR practitioners in implementing such solutions, as highlighted by Stephenson et al. \cite{Stephenson2022} in their meta-review.
The authors recognize the potential of behavioral biometrics-based solutions but criticize the complexity regarding their deployment.
This analysis aligns with our assessment, underscoring the need for production-ready versatile solutions that can be deployed without the necessity of training expertise or resources.

Our approach combines the pretrainability of feature-distance methods with the generalization capability of classification-learning models through the development of a similarity-learning model.
We apply Deep Metric Learning (DML)~\cite{musgraveMetricLearningReality2020}, a method that has shown impressive results in various fields like face recognition and item retrieval~\cite{kaya2019deep}.
Unlike classification-learning models which produce immediate classifications and require retraining for new classes, DML models produce mappings of their inputs into a high dimensional representation space.
Specifically, the idea is to group similar items together and separate dissimilar ones within this learned space, which is why this method is commonly referred to as \textit{similarity-learning}.
Since this also works for new classes (i.e., users not present during training), these models are pretrainable and efficient in onboarding new users, thus addressing the key limitations of current XR user identification methods.

Miller R. et al.~\cite{Miller2021} is the only previous work known to us to also employ a similarity-learning algorithm for user identification from motions, specifically utilizing a technique they refer to as ``Siamese Neural Networks" (SNN).
SNNs belong to the category of DML models that utilize tuple-based loss functions, considering pairs or triplets of input items.
Although the approach from Miller R. et al. conceptually resembles ours, it focuses on the SNNs' ability to manage device-specific noise.
Furthermore, their experimental design reserves only a single user for the test set, evaluating the model's performance against one unknown user (not included in the training set) in conjunction with known motion sequences from known users.
This approach does not answer the question regarding the SNN's capability to differentiate unknown from known users, or even just known from unknown motion sequences, leaving the extent of SNNs' generalization capabilities ambiguous.
Therefore, while their findings might indicate SNNs' ability to generalize across devices, the comprehensive potential of DML models for adapting to new users and achieving generalization across a wider array of tasks remains unexplored.

In contrast, our work takes a comprehensive view of the similarity-learning approach, conducting a thorough analysis to underscore its benefits over traditional classification-learning and feature-distance methods.
We not only incorporate the tuple-based loss function used by Miller R. et al. but also investigate other types of loss functions, such as proxy-based losses (detailed in \Cref{sec:loss_functions}).
Crucially, our experimental setup includes a comparison between the similarity-learning model and a classification-learning model, providing insights into their respective performance differences.
This comparative approach allows us to highlight the enhanced versatility and the possibility of using pretrained similarity-learning models, particularly in scenarios involving unpredictable user motions.
Moreover, we pay attention to setting up a sound train-validation-test split to avoid data leakage \cite{KapoorLeakageReproducibilityCrisis2022} from the pre-training process into evaluation.

%% file: sections/03_methods.tex
\section{Datasets}
\label{sec:dataset}

We use two datasets for our experiments: ``Who Is Alyx?" for pretraining and evaluation, as well as the dataset of Miller~R. et al.~\cite{Miller2022CombiningBiometrics} for evaluation.

\subsection{Who Is Alyx?}
We used the ``Who~Is~Alyx?'' dataset \cite{RackWhoAlyxnew2023} for pretraining and evaluating the similarity- and classification-learning models.
It includes the movement data of 63 users\footnote{the dataset now includes 71 users, but the recordings of the last 8 users were not available in time when we started our work for this article} playing the VR action game ``Half-Life:~Alyx'' with a HTC Vive Pro.
Each participant completed two gaming sessions on different days, averaging 46 minutes per session. 
The shortest three sessions ranged between 32 and 40 minutes. 

In the game, players assume the role of Alyx Vance and have to make their way through a futuristic dystopian world.
The player can navigate through the virtual world by either walking within the boundaries of the VR setup or using a teleportation mechanism.
Along the way, the player has to overcome a wide range of obstacles, including puzzles, enemies or simply finding the right way.
Altogether, this results in a wide range of different user motions.
The VR setup provided no option to sit down, so every participant was standing (or sometimes crouching) the whole time.
User motions were recorded with 90 frames per second.
Each frame consists of spatial (x,y,z) and rotational (quaternion: x,y,z,w) coordinates from the head mounted display (HMD) and both handheld controllers, totaling 21 features per frame.

We selected the ``Who~Is~Alyx?'' dataset for our experiments due to its suitability for training and evaluating a model with high versatility. 
This dataset is particularly advantageous due to its high variety of user motions. 
The range of physical activities in ``Half-Life:~Alyx'' --- from complex navigational maneuvers to intricate interactions with the game environment --- ensures exposure to a diverse set of motion patterns. 
This diversity is crucial for developing a model that can accurately discern user-specific motion signatures amidst a wide array of potential actions. 
Moreover, the dual-session structure of the dataset allowed for the examination of intra-user variability over time, adding another layer of complexity to the model's training and evaluation process. 
Thus, the ``Who~Is~Alyx?'' dataset provides an ideal context for training a robust, versatile model capable of handling the unpredictability and richness of human motions in VR environments.

\subsection{MR Dataset}
Besides the ``Who Is Alyx?'' dataset, we also use the dataset of Miller~R. et al.~\cite{Miller2022CombiningBiometrics}, which we refer to in the following as ``MR dataset''.
This dataset includes 41 users who perform the same specific ball-throwing action in two sessions.
In addition, the MR dataset was recorded with three different VR devices:
an HTC Vive, an HTC Cosmos, and an Oculus Quest.
The 41 users repeat the ball-throwing action 10 times for each device and each session, with each repetition being clipped to exactly 3 seconds.

Note that the MR dataset differs in several key aspects from the ``Who Is Alyx?" dataset: it not only includes an entirely different set of users but also a completely different type of very specific action, as well as different VR devices.
Hence, by using the MR dataset for testing, we can investigate how well the similarity-learning model works with VR users using different hardware and performing a motion that is not common in the training dataset.

\section{Methodology}
\label{sec:methodology}

In this section, we describe our methodology for pretraining the similarity-learning and using it for the subsequent identification task.
In our experiments, we also used a classification-learning model for baseline, which we describe in \Cref{sec:baseline-model} as part of our experimental setup.

\subsection{Deep Metric Learning}
We used Deep Metric Learning (DML) for our similarity-learning approach in identifying XR users based on motion data.
DML differs from conventional classification-learning by not directly classifying input data into predefined classes. 
Instead, it generates continuous vector representations, which are typically referred to as `embeddings'.
During training, DML models learn to minimize or maximize the distance between pairs (or groups) of embeddings, depending on whether they belong to the same class or not.
This also works for unseen classes, making this approach highly scalable and adaptable.
In our case, the model learned to detect distinct user motion profiles, thus placing incoming motion sequences of the same user together and sequences of different users farther apart.
These embeddings produced by the DML model allow to derive the identity in a similar way as previous feature-distance approaches:
feature-distance approaches compare the similarity between inputs by considering their distance (e.g., Euclidean) directly in feature space, while the similarity-learning approach considers the distance of inputs in the learned embedding space.

\subsection{Architecture}

For the DML approach, a wide range of deep learning architectures can be used, with recurrent neural networks (RNN) and convolutional neural networks (CNN) being among the most common choices.
In contrast to basic multilayer perceptrons or other basic machine learning models, RNNs and CNNs are able to work directly with sequential data (e.g., consecutive frames of motion data) without the need for any dimensionality reduction.
In our case, the choice of the architecture has been guided by the results from Rack et al. \cite{Rack2022ComparisonSequences} who compared three different types of RNNs (i.e., FRNN, GRU and LSTM) for a classification-learning approach to identify users. 
Here, gated recurrent units (GRU) and LSTMs work equally well and outperformed the other architectures.
For this work, we selected the GRU architecture, as it is a more parameter efficient variant of the LSTM.
We also made preliminary experiments with CNN architectures used by previous works \cite{Mathis2020,Miller2021,RackWhoAlyxnew2023}, which are interesting since they are trained much quicker, but we did not achieve results that could compete with the GRU models.
Since the main scope of our work was to evaluate the similarity-learning approach itself, we prioritized our resources to extensively investigating one specific architecture (i.e., the GRU) over comparing and analyzing multiple architectures.

We used the GRU implementation from the PyTorch library \cite{Paszke_PyTorch_An_Imperative_2019}, which exposes three hyperparameters we selected for tuning (see \Cref{sec:hyperparametersearch} and \Cref{tab:hyperparameter_search_space}): the `number of layers' determines the number of stacked recurrent layers, `layer size' refers to the size of the hidden state and `dropout' specifies the probability of dropping units during training to improve the model's robustness and prevent overfitting.

\subsection{Model Input}
\label{sec:input}

The GRU architecture requires input sequences of a fixed size.
We selected a sequence size of 500 frames, which represents 33 seconds given the constant 15 frames per second.
As each frame provides 18 features, the final model input shape is $500 \times 18$.
For sequences with more than 500 frames we sampled multiple subsequences by sliding a window of size 500 over the given sequence frame by frame.
For example, 1 minute contains $\SI{60}{seconds} \times \SI{15}{fps} = \SI{900}{frames}$; after taking the first window of 500 frames for the first subsequence, the window can be slid further frame by frame over the remaining 400 frames, resulting in a total of 400 subsequences.
Each subsequence was then independently given to the model, which in this example would result in 400 individual embeddings for this 1 minute sequence.

\subsection{Pretraining Phase}

In our methodology, the similarity-learning model is considered `pretrained' once it has been fully trained to map input sequences to an embedding space, reflective of user identities.
For each training step, a batch of subsequences from different users was sampled from the training dataset and fed through the model.
Here, each batch consisted of approximately equally many subsequences per training user.
The model outputs one embedding of dimensionality $n$ for each subsequence.
All embeddings in a batch were then given to the loss function to compare the distances between embeddings of the same and different users.
More on the choice of the network, the used loss functions, and the embedding dimensionality $n$ can be found in \Cref{sec:experimental_setup}, since these are variables that have been optimized during the hyperparameter search.

\subsection{Identification With Pretrained Model}
\label{sec:model_inference}

Once the similarity-learning DML model has been pretrained, it is able to map input sequences into the embedding space based on the identity of the user.
Before a new user can be identified, they have to be registered by collecting \textit{reference} sequences.
Registering new users in the context of classification-learning models would require a full training of the model.
However, we did not retrain the similarity-learning model to register new users.
Instead, we utilized it solely to turn the reference motion sequences into reference embeddings $R^u = R^u_1, R^u_2, \dots, R^u_{|R^u|}$ for each user $u \in U$.
This process of registering new users by generating and adding their reference embeddings into an existing collection of known users is significantly faster and can be done within seconds.
This is in stark contrast to training classification-learning models, which can extend to several hours or days on identical hardware and data. 

To identify users after registration, we retrieved their motion sequences, split them into subsequences and encoded them into embeddings $Q = Q_1, Q_2, \dots, Q_{|Q|}$.
These embeddings are called \textit{query} embeddings, since they will be used to query the reference collection.
If the model learned to effectively encode the user identity in the embedding space, it will produce query embeddings that are close (e.g., in terms of Euclidean distance) to the previously collected reference embeddings of the same user.
Querying the reference collection requires an efficient data structure that allows for fast nearest-neighbor search in large collections of high-dimensional vectors.
For our experiments, we use the FAISS library \cite{johnson2019billion} to perform the search in memory, but for more advanced use cases data structures like vector similarity databases such as Milvus~\cite{wang2021milvus} are appropriate.

Depending on its error rate, the DML model will not embed every input sequence perfectly, so the closest reference embedding for any given query embedding might actually belong to the wrong user.
We therefore did not only consider the first closest reference embedding, but the 50 closest (i.e., ``neighbouring'') reference embeddings to the query embedding.
Then, we checked the registered user of each of the 50 embeddings and picked the most common one.
This way we ended up with one predicted identity for each subsequence we had sampled from the users' motion sequence.
To receive a final prediction for the whole motion sequence of the user we performed a majority vote across the predictions of all subsequences.
Note, that we determined this value of 50 based on evaluations on the validation split during training.

\subsection{Data Preprocessing}

We followed the data preprocessing steps from Rack et al.~\cite{Rack2022ComparisonSequences} concerning resampling and data encoding.
The authors found that input sequences for neural networks should preferably cover a longer time period with fewer frames rather than a shorter time period with more frames.
Consequently, we resampled the original footage from 90 frames per second to 15 frames per second, which is in tune with these findings and also worked best during preliminary trials.
After resampling, we encoded the data, as this is an efficient way to remove noise and helps neural networks to generalize \cite{Rack2022ComparisonSequences}.
For this we referenced all coordinates to the head (i.e., body-relative; BR), which works in two steps.
First, for each frame, we rotated positions and orientations of the controllers by the yaw rotation of the HMD.
For example, if a user had turned by 90 degree, this procedure would rotate them back to face into the forward direction again.
Second, we subtracted the positional coordinates from the controllers by the corresponding coordinates from the head.
This placed the user back onto the scene's origin, even if they had changed their position.
Altogether, with this BR encoding, the same motion (e.g., hand waving) produces the same feature values independently from the user's position and orientation in the scene.

Next, we computed the deltas between each frame, effectively producing the velocities for each feature (i.e., body-relative velocity; BRV).
Then, we took this idea one step further and computed the acceleration for all features (i.e., body-relative acceleration; BRA) by taking the deltas between each BRV frame, as preliminary trials had shown that models perform even better with accelerations than with velocities.
Altogether, we ended up with 18 BRA encoded features: (pos-x, pos-y, pos-z, rot-x, rot-y, rot-z, rot-w) $\times$ (controllers left \& right) + (rot-x, rot-y, rot-z, rot-w) $\times$ (HMD).
Note, that for the HMD there remains only one quaternion for pitch and roll rotations, since position and yaw rotation had been used during re-referencing of the controllers and become obsolete as a consequence.
All of these computations have been done using the Motion Learning Toolbox \cite{MLT}.

\section{Experimental Setup}
\label{sec:experimental_setup}

For our experiments, we evaluated a similarity-learning model and compared its performance against a classification-learning model with the same architecture as baseline.
To test how well the pretrained similarity-learning solution works, we held back 27 test users who had not been used to train or tune our models (find more details about the data split in \Cref{sec:datasplit}).
We follow the terminology used by the literature \cite{Rogers2015,Miller2022CombiningBiometrics,Miller2022TemporalBiometrics}, which describes the data collected during on-boarding of users as \textit{enrollment} and the data collected for later identification (in another session) as \textit{use-time} data.
In the experiment, we used the first session of each test user playing ``Half-Life: Alyx'' as enrollment data.
This enrollment data was used as reference data for our pretrained similarity-learning model and as training data for the classification-learning baseline model (compare `enrollment' in \Cref{sec:datasplit}).
Then, we considered the second session as use-time data to evaluate how well both models can identify each user.

Our research targets the identification task within biometric systems, as defined by Jain et al. \cite{biometrics_intro}, which distinguishes between ``identification" and ``verification" functionalities.
Identification systems use the biometric input from the user to query a database with biometric templates of all known users.
Verification systems compare the biometric input only against the templates of the identity claimed by the user.
We opted for focusing on the broader scope of identification to demonstrate our solution's versatility, as verification would require a more precise definition of specific verification scenarios.

For the identification task, we payed attention to simulate a real-world identification system:
in such systems, the amount for both, enrollment data and use-time data, that can be captured per user depends on the individual use case and can be very different each time.
For example, in a virtual classroom it may be no problem to record students for half an hour while they interact with each other and the teacher (given appropriate consent), but for use cases that consider short meetings, presentations etc. only a few minutes might be available.
Hence, we limited the sequence length of enrollment and use-time data to different lengths and varied this limit in our experiments (e.g., identifying a user within 5 minutes of use-time data based on 10 minutes of enrollment data).
More specifically, we evaluated enrollment data of 1, 5, 10, 15, 20, 25, 30, and 35 minutes, as well as all available data for a user.
For use-time data, we tested windows of 1, 5, 10, 15, 20, 25, and 30 minutes for each user.
This gives us an impression of how well the tested models perform in different scenarios, allowing us do discuss possible application scenarios for each method.
Note, that each of these windows is longer than the selected model input of 500 frames (33 seconds).
Hence, in each of these cases we sample subsequences as explained in \Cref{sec:input} and derive a final prediction for each window as explained in \Cref{sec:model_inference}.
In the following we describe the different aspects of our experimental setup in more detail.

\subsection{Loss Functions}
\label{sec:loss_functions}

Machine learning models are trained by using a loss function that evaluates the quality of the model output during each training iteration, which is in turn used by the optimizer to update the model's parameters.
In the context of training similarity-learning models, various loss functions can be utilized to evaluate the quality of the model output and guide the optimization process.
We explored five common options in our similarity-learning approach, which can be categorized into tuple-based losses and proxy-based losses.

Contrastive Loss~\cite{hadsellDimensionalityReductionLearning2006}, Triplet-Margin Loss~\cite{weinbergerDistanceMetricLearning2006}, and Multi-Similarity Loss~\cite{wangMultiSimilarityLossGeneral2019} are tuple based.
These losses use two or more inputs and define their loss by relating their embeddings to each other.
If both embeddings belong to the same class/user, the distance between them should be minimized.
If they come from different classes, the distance should be at least a specified minimal distance.

ArcFace Loss~\cite{dengArcFaceAdditiveAngular2019}, and Normalized Softmax Loss~\cite{liuSphereFaceDeepHypersphere2018,wangNormFaceL2Hypersphere2017,zhaiClassificationStrongBaseline2019} are proxy-based losses that define and optimize representative vectors for each class, so-called proxies.
Then, the distance between each example and its class proxy should be minimized while the distance to other proxies should be maximized.

While tuple-based losses usually perform better since they compare examples directly to each other, proxy-based losses are computationally more efficient, since they only relate examples with their proxies instead of having to regard all possible tuples in a dataset.
We thus experimented with both kinds of losses to estimate the most suitable loss function for this task.

Each loss can be configured and fine-tuned with its own set of parameters, which are listed in \Cref{tab:losses_search_space}.
To determine appropriate configurations, we performed an individual hyperparameter search for each loss, as explained in more detail in \Cref{sec:hyperparametersearch}.

\subsection{Data Split: Train, Validation, and Test}
\label{sec:datasplit}

\begin{figure}[t]
    \centering
    \includegraphics[width=\linewidth]{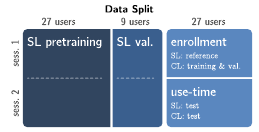}
    \caption{Data split of the 63 users for the similarity-learning (SL) and classification-learning (CL) models; the SL model was pretrained and validated with 36 users; the remaining 27 users were used for evaluation of SL and CL.}
    \label{fig:data_split}
\end{figure}

We adopted a dataset split strategy that is suitable to substantiate our claim of `pretrainability', summarized in \Cref{fig:data_split}.
Following common practice in similarity-learning research, we used approximately half of the dataset classes for training and the other half for testing, which offers a good balance between training dataset size and test set variety~\cite{musgraveMetricLearningReality2020}.
We therefore split the 63 users from the ``Who~Is~Alyx?'' dataset into three different subsets: 27 training users, 9 validation users, and 27 testing users.

During the pretraining phase, the training set was used for the actual training, while the validation set helped to monitor this process, to compare different training runs and select the final model hyperparameters.
The test dataset was reserved solely for the final evaluation.
We selected the first session of the test set as enrollment data, serving as the reference sequences for each user. 
Consequently, we selected the second session as use-time data, serving as query embeddings, with the goal of identifying the performing user from the test user group.

By employing this approach, we were able to test the pretrained similarity-learning model on a set of users who were not part of the pretraining phase.
This setup effectively demonstrates the model's capacity to generalize and accurately identify users who were previously unknown to it.

\subsection{Classification-Learning Baseline Model}
\label{sec:baseline-model}

To compare our proposed similarity-learning method to the prevailing state of the art from the literature, we also evaluated a classification-learning model as a baseline.
For a fair comparison, we used the same architecture (i.e., a GRU architecture), so the only conceptual differences between the two models were the type of output (i.e., classification vs. embedding) and the training procedure (i.e., Categorical Cross Entropy vs. our tested DML loss functions).
We also used the same data to evaluate the baseline model.
Here, the enrollment data was used to train the model and the use-time data to test it.
As with the similarity-learning model, we varied the amount of available data during both, enrollment and use-time.
From this data, input subsequences are generated and preprocessed in the same way as for the similarity-learning method.
The classification-learning model predicts one user for each subsequence.
To make one prediction for a longer sequence, we performed a majority vote over all subsequence predictions in the complete sequence.

\subsection{Evaluation Metric}
\label{sec:sequence_accuracy}

Given a certain period of enrollment data $t_\text{enr}$ per user (in minutes) and use-time data $t_{use}$ (in minutes), we computed the prediction accuracy of an identification method \acc{t_\text{enr}}{t_{use}}, i.e., the proportion of sequences with length $t_{use}$ that were correctly identified as the corresponding user, for which we had $t_\text{enr}$ minutes of enrollment data in our user data collection.
Overall, we extracted all time windows of length $t_\text{use}$ in a sliding window approach in one second steps from the use-case data (session two).
For instance, an accuracy of \acc{5}{10} = 0.7 indicates that 70\% of all 10-minute sequences were accurately identified with 5 minutes of enrollment data per user.
In general, we applied macro-averaging to report the average accuracy per class/user.
This way we accounted for eventual class imbalances in the test set, even though the whole dataset was fairly balanced, as there was about the same amount of footage for each user.

We also computed other metrics like the Matthews Correlation Coefficient or Cohen's Kappa, but they consistently mirrored the macro-averaged accuracy, offering no additional insights.
Consequently, for simplicity and clarity we did not report additional metrics.

\subsection{Hyperparameter Search}
\label{sec:hyperparametersearch}
\begin{table}[]
\caption{Hyperparameter search space for the similarity-learning model and the classification-learning baseline model.}
\label{tab:hyperparameter_search_space}
\centering
\begin{tabular}{lll}
\toprule
\textbf{Parameter}       & \textbf{Search Space} \\
\midrule
\textbf{GRU layers}      & [1, 4]                \\
\textbf{GRU layer size}  & [100, 500]            \\
\textbf{GRU dropout}     & [0, 0.4]              \\
\textbf{learning rate}   & [$10^{-6}$, 0.1]      \\
\textbf{loss (similarity-learning only)} & see \autoref{tab:losses_search_space} \\
\textbf{embedding size $n$ (similarity-learning only)} & [32, 320]   \\
\bottomrule
\end{tabular}
\end{table}
\begin{table}
\caption{Hyperparameter search spaces for DML loss functions.}
\label{tab:losses_search_space}
\centering
\begin{tabularx}{\linewidth}{lll}
\toprule
\textbf{Loss}                              & \textbf{Parameter} & \textbf{Search Space} \\
\midrule
\multirow[t]{2}{*}{\textbf{Contrastive}}   & pos. margin        & [0, 0.3]            \\
                                           & neg. margin        & [0.3, 1]             \\
\midrule
\textbf{Triplet-Margin}                    & margin             & [0.01, 0.5]           \\
\midrule
\multirow[t]{3}{*}{\textbf{Multi-Similarity}} & alpha              & [0.01, 20]         \\
                                           & beta               & [20, 80]              \\
                                           & base               & [0, 3]                \\
\midrule
\multirow[t]{3}{*}{\textbf{ArcFace}}       & regularizer        & [$10^{-6}$, 0.1]       \\
                                           & margin             & [1, 20]                  \\
                                           & scale              & [1, 500]                 \\
\midrule
\textbf{Normalized Softmax}                & temperature        & [$10^{-5}$, 0.01]   \\
\bottomrule
\end{tabularx}
\end{table}

Deep learning architectures, such as the GRU used in this work, typically come with several configuration options that have a substantial impact on their performance.
These \emph{hyperparameters} cannot be known beforehand but need to be determined via trial and error \cite{Wolpert1997NoOptimization}.
We use the service Weights \& Biases \cite{wandb} to monitor and coordinate the hyperparameter searches for the similarity-learning and classification-learning models, for which the investigated search spaces can be found in \autoref{tab:hyperparameter_search_space} and \autoref{tab:losses_search_space}.
This service provided a Bayesian controlled search to suggest new configurations during the hyperparameter search by basing its suggestions on a defined target metric of previous runs.

For the similarity-learning models we used the accuracy given all enrollment data and 5 minutes of use-time data (\acc{all}{5}) on the nine validation users as target metric.
The classification-learning baseline models were trained on session one of the test users, of which we held back the last 5 minutes for validation and considered the mean sequence accuracy as target metric.
Altogether, we performed five separate hyperparameter searches for the similarity-learning models (one for each loss) and one for the classification-learning model.

\subsection{Training \& Implementation}

We used PyTorch Lightning \cite{Falcon_PyTorch_Lightning_2019} together with the metric learning library from Musgrave et al. \cite{Musgrave2020}.
For optimization, we used Adam~\cite{kingma2015adam} and either one of the DML specific losses (\Cref{sec:loss_functions}) for the similarity-learning approach, or the Categorical Cross Entropy for the classification-learning model.
We trained each model for at least 100 epochs until the performance on the target metric stagnated or declined.
During training we saved the best models based on their validation performances.
Each training was performed with 8 CPU units, one GPU (either NVIDIA GTX 1080Ti, RTX 2070 Ti, or RTX 2080Ti), and 20 Gb RAM on the computing cluster of our institute.

%% file: sections/04_results.tex
\section{Results}

Altogether we trained at least 140 models with different architecture configurations for each of the five loss functions, totaling in over 850 runs for hyperparameter searches of the similarity-learning models.
Additionally, we perform 100 training runs for the classification-learning baseline model.
The results reported in this section are based on the hyperparameter configurations that yielded the highest validation accuracy for either model, as detailed in \autoref{tab:final_model_configurations}.

\begin{table}[]
\caption{Final configuration for baseline and DML model after hyperparameter search.}
\label{tab:final_model_configurations}
\centering
\begin{tabular}{@{}lp{2cm}p{3cm}@{}}
\toprule
                        & \textbf{Classification-Based Model} & \textbf{Embedding-Based Model} \\
\midrule
\textbf{GRU layers}     & 4                       & 3                  \\
\textbf{GRU layer size} & 200                     & 450                \\
\textbf{GRU dropout}    & 0.19                    & 0.28               \\
\textbf{learning rate}  & $7\times 10^{-4}$       & $2\times 10^{-5}$  \\
\textbf{loss}           & Cross Entropy           & ArcFace (reg.: $9\times 10^{-5}$, margin: 3.5, scale: 211)     \\
\textbf{embedding size $n$} & ---                 & 192                \\
\midrule
\textbf{trainable parameters} & 3,161,676 & 861,027 \\
\bottomrule
\end{tabular}
\end{table}

\begin{figure*}[hbt]
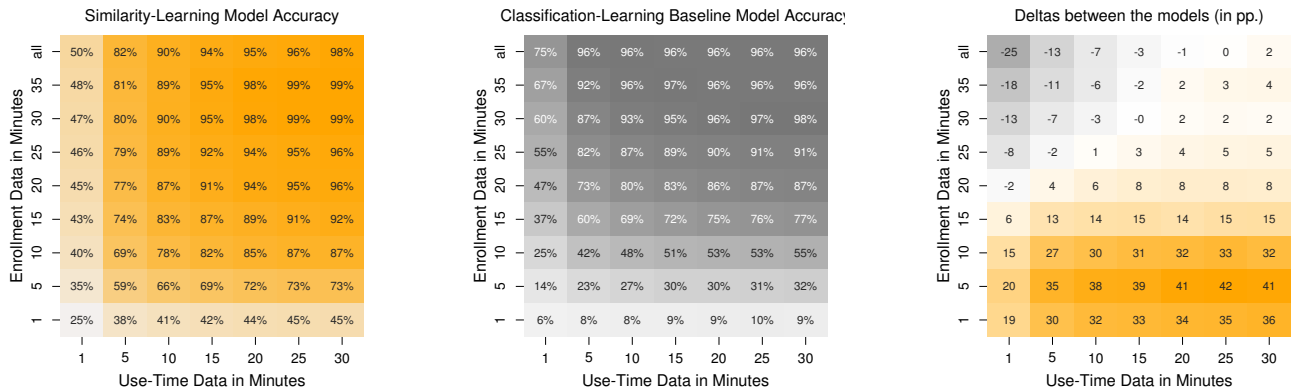

    \subfloat[Accuracies of our proposed similarity-learning model.]{
        \resizebox{0.3\textwidth}{!}{
            \input{results/generated/dml_model_sequence_accuracy.pgf}
        }
        \label{fig:dml_model_sequence_accuracy}
    }
    \hfill
    \subfloat[Accuracies of classification-learning baseline model.]{
        \resizebox{0.3\textwidth}{!}{
            \input{results/generated/baseline_model_sequence_accuracy.pgf}
        }
        \label{fig:baseline_model_sequence_accuracy}
    }
    \hfill
    \subfloat[Deltas between similarity-learning model (\autoref{fig:dml_model_sequence_accuracy}) and classification-learning baseline model (\autoref{fig:baseline_model_sequence_accuracy}).]{
        \resizebox{0.3\textwidth}{!}{
            \input{results/generated/delta_seq_accuracies.pgf}
        }
        \label{fig:delta_seq_accuracies}
    }
    
    \caption{Results for our experiments for the proposed similarity-learning method, as well as the baseline classification-learning model. We also visualize the difference between both models to highlight performance gaps.}
    \label{fig:results}
\end{figure*}

\subsection{Evaluation}

\begin{figure}[hbt]
  \input{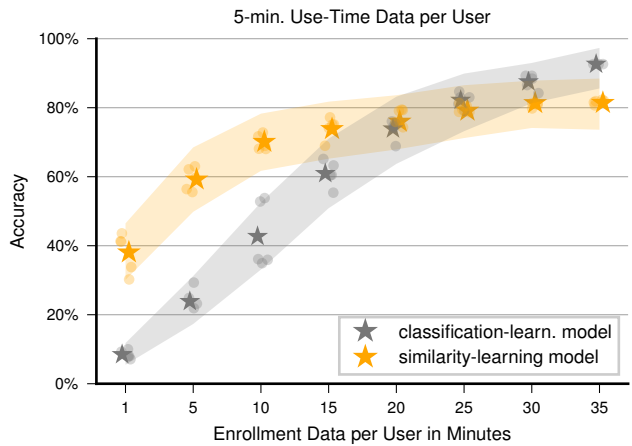}
  \caption{Detailed view of the 5-minute columns from \Cref{fig:dml_model_sequence_accuracy}; the enrollment procedure is repeated 5 times for each model and duration, each repetition denoted with a circle, stars denote the means; shaded areas indicate 95\% confidence intervals (bootstrapped).}
  \label{fig:train_time_comparison}
\end{figure}

\Cref{fig:results} shows the results for our experiments.
\Cref{fig:dml_model_sequence_accuracy} displays the accuracies for our proposed similarity-learning approach for different enrollment and use-time sequences, while \Cref{fig:baseline_model_sequence_accuracy} consists of the corresponding results for the classification-learning baseline method.
For any other enrollment-length than ``all'' we selected for each user a random starting point in the first session.
To compensate for stochastic effects that this procedure might have, we repeated the evaluation of each enrollment-length five times, and report the averaged results.
This can be seen in \Cref{fig:train_time_comparison}, where small circles mark the accuracy of the individual evaluations, and stars mark the averaged result.
We also report the differences between both models in \Cref{fig:delta_seq_accuracies} for an easier comparison.

Several general trends emerge from our analysis. 
Firstly, it is evident that both the similarity-learning and classification-learning models greatly outperform a random guess baseline, which would have an expected accuracy of approximately $\frac{1}{27}\approx 3.7\%$. 
Furthermore, there is a clear correlation between the amount of enrollment and use-time data and the overall identification accuracy for both models: as either dimension of data increased, so did performance. 
Notably, increasing the enrollment data tended to yield larger gains in accuracy than proportional increases in use-time data. 
This suggests that having a large repertoire of reference motions for each user is preferable to having more footage during use-time.
Lastly, when all data for enrollment and use-time was used, both models approached perfect accuracy.

\Cref{fig:delta_seq_accuracies} and \Cref{fig:train_time_comparison} both highlight a key strength of the similarity-learning model: its superior performance with limited amounts of reference motion data. 
This can be attributed to the pretraining phase, during which the model has already learned general human motion characteristics. 
In contrast, the classification-learning model apparently lacked sufficient data at lower enrollment levels to effectively learn how humans move and how to discern individual motion profiles.
However, as the amount of enrollment data grew, the classification-learning model's performance improved and eventually surpassed the similarity-learning model. 
With enough data, the classification-learning model was able to specialize to the specific set of users in its training set.
In essence, the pretrained nature of the similarity-learning model equips it with a foundational understanding of human motion that is beneficial when data is scarce. 
Yet, with sufficient data, the classification-learning model can leverage its ability to focus on distinctive characteristics of its training set, thereby gaining an edge in accuracy.

To statistically validate the differences observed in model performances, we utilized the Wilcoxon Signed Rank Test \cite{wilcoxon_45,wilcoxon_92}. 
This non-parametric test assesses whether two related samples come from the same distribution, specifically useful for comparing two sets of matched pairs. 
In our case, it is applied to determine if the differences in accuracy between the similarity-learning and classification-learning models are statistically significant or could have occurred by random chance.
The outcomes of the test confirmed the visual indications from \Cref{fig:train_time_comparison}: 
for the comparisons \acc{20}{5}, \acc{25}{5}, and \acc{30}{5}, the differences in performance were not statistically significant, implying that the two models performed comparably under these conditions. 
However, for all other comparisons, the p-values fell below the 0.02 threshold. 
This suggests a high level of confidence that the observed performance disparities, especially in scenarios with less enrollment data, are statistically significant and not due to random variation.

\subsection{Analysis}

In the following, we perform deeper analyses to investigate the robustness and potentials of our proposed similarity-learning approach.

\subsubsection{Is the model robust to different seeds?}

The performance of deep learning models not only depends on the selected hyperparameter configurations, but is also influenced by random factors during training.
The seed used to initialize the model's weights with random values or the used GPU had impact on the course of the training and the resulting trained model.
Additionally, the split of users into train, validation, and test sets could also have had an impact on the performance.
To gain insight into how sensitive our proposed similarity-learning method is to these factors, we retrained the model with the best hyperparameters 25 times: 1) with 5 different seeds and, 2) for each seed, with 5 different user data split assignments.
Given the \acc{5}{1}, we found that the proposed approach is relatively robust against different seeds:
within the same data split assignments, the accuracy only varied within 2.4 percentage points.
For different data split assignments, the differences in the test scores varied within 7 percentage points.
This gets smaller the more enrollment data we use, for \acc{all}{1} different data split assignments vary only by 2 percentage points.

\subsubsection{How does the number of training users affect the model?}

\begin{figure}[hbt]
  \input{results/generated/number_of_training_subjects.pgf}
  \caption{Training of the similarity-learning model with fewer subjects in training data set; reported is \acc{all}{1} on the test dataset (with all 27 test subjects in each instance).}
  \label{fig:number_of_training_subjects}
\end{figure}
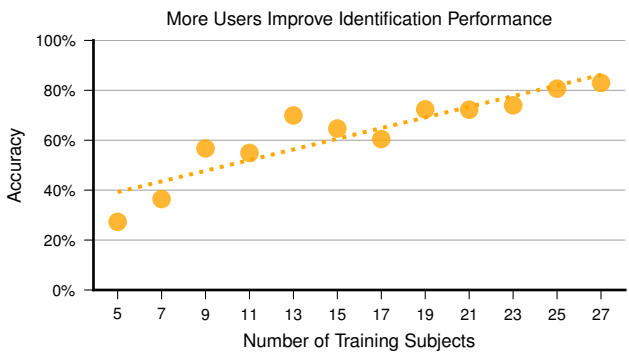

In our experiments, we used the data of 27 users to train the proposed similarity-learning method.
To see how the performance depends on the number of users in the training set, we retrained the model with a varying number of training subjects while letting the validation and test splits remain unchanged.
For this, we again used the same hyperparameters as before.
The results are reported in \Cref{fig:number_of_training_subjects}.
As expected, a larger set of training users increased the performance of the similarity-learning model.
We also observed that the performance is not saturated for the maximum available training users for our dataset.
This suggests that additional users during training can substantially improve current results, which is a common observation in related applications such as face recognition, where production-grade models are typically trained on millions of subjects~\cite{zhu2022webface260m}.
This motivates the collection of larger datasets for the task of VR user identification.

\subsubsection{Does the pretrained model generalize to the dataset from Miller~R. et~al.?}

As a final analysis, we investigated the generalization ability of our proposed similarity-learning method in terms of movements that are unusual for the used training dataset and how the model can generalize across different VR systems.
For this analysis we used the 41 users from the MR dataset for testing.
This allowed us to re-purpose the test-split from ``Who is Alyx?'' described in \Cref{sec:datasplit} and add it to the pretraining phase of the similarity-learning model.
Consequently, we used 50 users for training and 13 for validation from ``Who is Alyx?'', and all 41 users from the MR dataset for testing.

With this setup, we performed a hyperparameter search (75 runs) to find an appropriate configuration for the similarity-learning model.
Since the 3 second sequences from the MR dataset are much shorter than the 33 seconds sequences we had worked with before, we reconfigured the pretraining phase to process inputs of 135 frames with 45 fps.
Then, we used session one of the MR dataset for enrollment and identify the users from their use-time data in session two.
Since there were only 10 embeddings per user and device in the reference database after enrollment, we adjusted the number of considered nearest neighbor reference embeddings (as explained in \Cref{sec:model_inference}) from 50 to 10 ––– otherwise we would inevitably consider wrong users.
For evaluation, we first tested the model on each device separately (Cosmos, Quest, and Vive) for enrollment and use-time, and then combined the data of all devices (all).
Since we only have 10 repetitions per user and device as enrollment data, we did not vary the amount of enrollment data in this analysis.
Similarly, we did not vary the amount of use-time data, since we already used the entire sequences (i.e., 3 seconds) and did not sample subsequences.

We cannot directly compare our results to the model by Miller R. et al., who use a different evaluation scenario and metrics for their dataset.
Instead, we use the classification-learning baseline model again for comparison.
We noticed that training the classification-learning model on the data of only one device yields inferior results, presumably because the training data is too small.
Therefore, we trained the model on all devices of the first session, taking the first 7 repetitions for training and the remaining 3 for validation.
Then, after a hyperparameter search (230 runs), we used the resulting configuration to identify the users in session two.

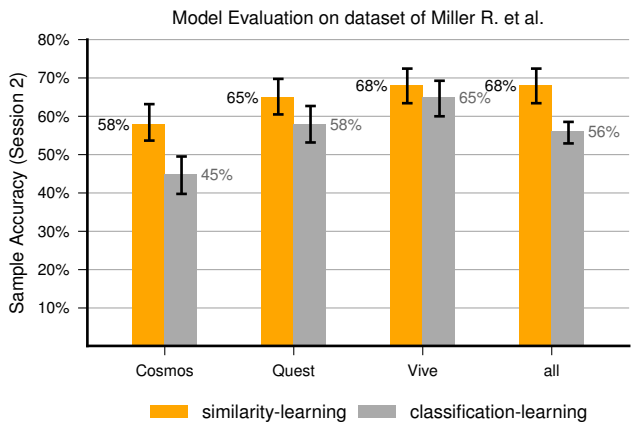
\begin{figure}
    \centering
    \input{results/generated/miller_evaluation.pgf}
    \caption{Evaluation of the similarity-learning and classification-learning models on the MR dataset; from the 41 subjects we use session one for enrollment, session two for use-time evaluation (10 sequences à 3 seconds per user per session); error bars indicate 95\% confidence intervals (bootstrapped).}
    \label{fig:miller_evaluation}
\end{figure}

The results from our extended analysis are detailed in \Cref{fig:miller_evaluation}. 
Despite the increased number of users (41 compared to 27 in our previous experiment) and limited enrollment and use-time data, both models significantly outperformed the random guess accuracy of approximately 2.5\%. 
Notably, the similarity-learning approach demonstrated robust performance across different VR devices, with particularly strong results on the HTC Vive dataset. 
This is likely due to the ``Who Is Alyx?" dataset being recorded on a HTC Vive Pro, suggesting the two devices share tracking characteristics.
Performance does dip slightly with enrollment and use-time data sourced from other VR devices, yet the similarity-learning model still maintained an edge over the classification-learning baseline. 
When leveraging data from all devices combined, the similarity-learning model's performance exceeds that of the classification-learning baseline by a notable twelve percentage points. 
This reinforces our previous hypothesis that pretrained similarity-learning models are more adept at working with smaller sets of enrollment data.

Ultimately, this analysis allows three observations: 1) the similarity-learning model can be pretrained on one dataset, and then be used without retraining to identify new users from a different dataset, 2) our model outperforms a classification-learning model that has specifically been trained on the MR dataset, and 3) we can verify the robust generalization across various VR devices observed by Miller R. et al. \cite{Miller2021}.

%% file: results/generated/number_of_training_subjects.pgf
\begingroup%
\makeatletter%
\begin{pgfpicture}%
\pgfpathrectangle{\pgfpointorigin}{\pgfqpoint{3.477004in}{2.000000in}}%
\pgfusepath{use as bounding box, clip}%
\begin{pgfscope}%
\pgfsetbuttcap%
\pgfsetmiterjoin%
\definecolor{currentfill}{rgb}{1.000000,1.000000,1.000000}%
\pgfsetfillcolor{currentfill}%
\pgfsetlinewidth{0.000000pt}%
\definecolor{currentstroke}{rgb}{1.000000,1.000000,1.000000}%
\pgfsetstrokecolor{currentstroke}%
\pgfsetdash{}{0pt}%
\pgfpathmoveto{\pgfqpoint{0.000000in}{0.000000in}}%
\pgfpathlineto{\pgfqpoint{3.477004in}{0.000000in}}%
\pgfpathlineto{\pgfqpoint{3.477004in}{2.000000in}}%
\pgfpathlineto{\pgfqpoint{0.000000in}{2.000000in}}%
\pgfpathlineto{\pgfqpoint{0.000000in}{0.000000in}}%
\pgfpathclose%
\pgfusepath{fill}%
\end{pgfscope}%
\begin{pgfscope}%
\pgfsetbuttcap%
\pgfsetmiterjoin%
\definecolor{currentfill}{rgb}{1.000000,1.000000,1.000000}%
\pgfsetfillcolor{currentfill}%
\pgfsetlinewidth{0.000000pt}%
\definecolor{currentstroke}{rgb}{0.000000,0.000000,0.000000}%
\pgfsetstrokecolor{currentstroke}%
\pgfsetstrokeopacity{0.000000}%
\pgfsetdash{}{0pt}%
\pgfpathmoveto{\pgfqpoint{0.589028in}{0.431111in}}%
\pgfpathlineto{\pgfqpoint{3.372004in}{0.431111in}}%
\pgfpathlineto{\pgfqpoint{3.372004in}{1.738333in}}%
\pgfpathlineto{\pgfqpoint{0.589028in}{1.738333in}}%
\pgfpathlineto{\pgfqpoint{0.589028in}{0.431111in}}%
\pgfpathclose%
\pgfusepath{fill}%
\end{pgfscope}%
\begin{pgfscope}%
\pgfsetbuttcap%
\pgfsetroundjoin%
\definecolor{currentfill}{rgb}{0.000000,0.000000,0.000000}%
\pgfsetfillcolor{currentfill}%
\pgfsetlinewidth{0.401500pt}%
\definecolor{currentstroke}{rgb}{0.000000,0.000000,0.000000}%
\pgfsetstrokecolor{currentstroke}%
\pgfsetdash{}{0pt}%
\pgfsys@defobject{currentmarker}{\pgfqpoint{0.000000in}{-0.048611in}}{\pgfqpoint{0.000000in}{0.000000in}}{%
\pgfpathmoveto{\pgfqpoint{0.000000in}{0.000000in}}%
\pgfpathlineto{\pgfqpoint{0.000000in}{-0.048611in}}%
\pgfusepath{stroke,fill}%
}%
\begin{pgfscope}%
\pgfsys@transformshift{0.715527in}{0.431111in}%
\pgfsys@useobject{currentmarker}{}%
\end{pgfscope}%
\end{pgfscope}%
\begin{pgfscope}%
\definecolor{textcolor}{rgb}{0.000000,0.000000,0.000000}%
\pgfsetstrokecolor{textcolor}%
\pgfsetfillcolor{textcolor}%
\pgftext[x=0.715527in,y=0.333889in,,top]{\color{textcolor}\sffamily\fontsize{6.000000}{7.200000}\selectfont 5}%
\end{pgfscope}%
\begin{pgfscope}%
\pgfsetbuttcap%
\pgfsetroundjoin%
\definecolor{currentfill}{rgb}{0.000000,0.000000,0.000000}%
\pgfsetfillcolor{currentfill}%
\pgfsetlinewidth{0.401500pt}%
\definecolor{currentstroke}{rgb}{0.000000,0.000000,0.000000}%
\pgfsetstrokecolor{currentstroke}%
\pgfsetdash{}{0pt}%
\pgfsys@defobject{currentmarker}{\pgfqpoint{0.000000in}{-0.048611in}}{\pgfqpoint{0.000000in}{0.000000in}}{%
\pgfpathmoveto{\pgfqpoint{0.000000in}{0.000000in}}%
\pgfpathlineto{\pgfqpoint{0.000000in}{-0.048611in}}%
\pgfusepath{stroke,fill}%
}%
\begin{pgfscope}%
\pgfsys@transformshift{0.945525in}{0.431111in}%
\pgfsys@useobject{currentmarker}{}%
\end{pgfscope}%
\end{pgfscope}%
\begin{pgfscope}%
\definecolor{textcolor}{rgb}{0.000000,0.000000,0.000000}%
\pgfsetstrokecolor{textcolor}%
\pgfsetfillcolor{textcolor}%
\pgftext[x=0.945525in,y=0.333889in,,top]{\color{textcolor}\sffamily\fontsize{6.000000}{7.200000}\selectfont 7}%
\end{pgfscope}%
\begin{pgfscope}%
\pgfsetbuttcap%
\pgfsetroundjoin%
\definecolor{currentfill}{rgb}{0.000000,0.000000,0.000000}%
\pgfsetfillcolor{currentfill}%
\pgfsetlinewidth{0.401500pt}%
\definecolor{currentstroke}{rgb}{0.000000,0.000000,0.000000}%
\pgfsetstrokecolor{currentstroke}%
\pgfsetdash{}{0pt}%
\pgfsys@defobject{currentmarker}{\pgfqpoint{0.000000in}{-0.048611in}}{\pgfqpoint{0.000000in}{0.000000in}}{%
\pgfpathmoveto{\pgfqpoint{0.000000in}{0.000000in}}%
\pgfpathlineto{\pgfqpoint{0.000000in}{-0.048611in}}%
\pgfusepath{stroke,fill}%
}%
\begin{pgfscope}%
\pgfsys@transformshift{1.175523in}{0.431111in}%
\pgfsys@useobject{currentmarker}{}%
\end{pgfscope}%
\end{pgfscope}%
\begin{pgfscope}%
\definecolor{textcolor}{rgb}{0.000000,0.000000,0.000000}%
\pgfsetstrokecolor{textcolor}%
\pgfsetfillcolor{textcolor}%
\pgftext[x=1.175523in,y=0.333889in,,top]{\color{textcolor}\sffamily\fontsize{6.000000}{7.200000}\selectfont 9}%
\end{pgfscope}%
\begin{pgfscope}%
\pgfsetbuttcap%
\pgfsetroundjoin%
\definecolor{currentfill}{rgb}{0.000000,0.000000,0.000000}%
\pgfsetfillcolor{currentfill}%
\pgfsetlinewidth{0.401500pt}%
\definecolor{currentstroke}{rgb}{0.000000,0.000000,0.000000}%
\pgfsetstrokecolor{currentstroke}%
\pgfsetdash{}{0pt}%
\pgfsys@defobject{currentmarker}{\pgfqpoint{0.000000in}{-0.048611in}}{\pgfqpoint{0.000000in}{0.000000in}}{%
\pgfpathmoveto{\pgfqpoint{0.000000in}{0.000000in}}%
\pgfpathlineto{\pgfqpoint{0.000000in}{-0.048611in}}%
\pgfusepath{stroke,fill}%
}%
\begin{pgfscope}%
\pgfsys@transformshift{1.405521in}{0.431111in}%
\pgfsys@useobject{currentmarker}{}%
\end{pgfscope}%
\end{pgfscope}%
\begin{pgfscope}%
\definecolor{textcolor}{rgb}{0.000000,0.000000,0.000000}%
\pgfsetstrokecolor{textcolor}%
\pgfsetfillcolor{textcolor}%
\pgftext[x=1.405521in,y=0.333889in,,top]{\color{textcolor}\sffamily\fontsize{6.000000}{7.200000}\selectfont 11}%
\end{pgfscope}%
\begin{pgfscope}%
\pgfsetbuttcap%
\pgfsetroundjoin%
\definecolor{currentfill}{rgb}{0.000000,0.000000,0.000000}%
\pgfsetfillcolor{currentfill}%
\pgfsetlinewidth{0.401500pt}%
\definecolor{currentstroke}{rgb}{0.000000,0.000000,0.000000}%
\pgfsetstrokecolor{currentstroke}%
\pgfsetdash{}{0pt}%
\pgfsys@defobject{currentmarker}{\pgfqpoint{0.000000in}{-0.048611in}}{\pgfqpoint{0.000000in}{0.000000in}}{%
\pgfpathmoveto{\pgfqpoint{0.000000in}{0.000000in}}%
\pgfpathlineto{\pgfqpoint{0.000000in}{-0.048611in}}%
\pgfusepath{stroke,fill}%
}%
\begin{pgfscope}%
\pgfsys@transformshift{1.635519in}{0.431111in}%
\pgfsys@useobject{currentmarker}{}%
\end{pgfscope}%
\end{pgfscope}%
\begin{pgfscope}%
\definecolor{textcolor}{rgb}{0.000000,0.000000,0.000000}%
\pgfsetstrokecolor{textcolor}%
\pgfsetfillcolor{textcolor}%
\pgftext[x=1.635519in,y=0.333889in,,top]{\color{textcolor}\sffamily\fontsize{6.000000}{7.200000}\selectfont 13}%
\end{pgfscope}%
\begin{pgfscope}%
\pgfsetbuttcap%
\pgfsetroundjoin%
\definecolor{currentfill}{rgb}{0.000000,0.000000,0.000000}%
\pgfsetfillcolor{currentfill}%
\pgfsetlinewidth{0.401500pt}%
\definecolor{currentstroke}{rgb}{0.000000,0.000000,0.000000}%
\pgfsetstrokecolor{currentstroke}%
\pgfsetdash{}{0pt}%
\pgfsys@defobject{currentmarker}{\pgfqpoint{0.000000in}{-0.048611in}}{\pgfqpoint{0.000000in}{0.000000in}}{%
\pgfpathmoveto{\pgfqpoint{0.000000in}{0.000000in}}%
\pgfpathlineto{\pgfqpoint{0.000000in}{-0.048611in}}%
\pgfusepath{stroke,fill}%
}%
\begin{pgfscope}%
\pgfsys@transformshift{1.865517in}{0.431111in}%
\pgfsys@useobject{currentmarker}{}%
\end{pgfscope}%
\end{pgfscope}%
\begin{pgfscope}%
\definecolor{textcolor}{rgb}{0.000000,0.000000,0.000000}%
\pgfsetstrokecolor{textcolor}%
\pgfsetfillcolor{textcolor}%
\pgftext[x=1.865517in,y=0.333889in,,top]{\color{textcolor}\sffamily\fontsize{6.000000}{7.200000}\selectfont 15}%
\end{pgfscope}%
\begin{pgfscope}%
\pgfsetbuttcap%
\pgfsetroundjoin%
\definecolor{currentfill}{rgb}{0.000000,0.000000,0.000000}%
\pgfsetfillcolor{currentfill}%
\pgfsetlinewidth{0.401500pt}%
\definecolor{currentstroke}{rgb}{0.000000,0.000000,0.000000}%
\pgfsetstrokecolor{currentstroke}%
\pgfsetdash{}{0pt}%
\pgfsys@defobject{currentmarker}{\pgfqpoint{0.000000in}{-0.048611in}}{\pgfqpoint{0.000000in}{0.000000in}}{%
\pgfpathmoveto{\pgfqpoint{0.000000in}{0.000000in}}%
\pgfpathlineto{\pgfqpoint{0.000000in}{-0.048611in}}%
\pgfusepath{stroke,fill}%
}%
\begin{pgfscope}%
\pgfsys@transformshift{2.095515in}{0.431111in}%
\pgfsys@useobject{currentmarker}{}%
\end{pgfscope}%
\end{pgfscope}%
\begin{pgfscope}%
\definecolor{textcolor}{rgb}{0.000000,0.000000,0.000000}%
\pgfsetstrokecolor{textcolor}%
\pgfsetfillcolor{textcolor}%
\pgftext[x=2.095515in,y=0.333889in,,top]{\color{textcolor}\sffamily\fontsize{6.000000}{7.200000}\selectfont 17}%
\end{pgfscope}%
\begin{pgfscope}%
\pgfsetbuttcap%
\pgfsetroundjoin%
\definecolor{currentfill}{rgb}{0.000000,0.000000,0.000000}%
\pgfsetfillcolor{currentfill}%
\pgfsetlinewidth{0.401500pt}%
\definecolor{currentstroke}{rgb}{0.000000,0.000000,0.000000}%
\pgfsetstrokecolor{currentstroke}%
\pgfsetdash{}{0pt}%
\pgfsys@defobject{currentmarker}{\pgfqpoint{0.000000in}{-0.048611in}}{\pgfqpoint{0.000000in}{0.000000in}}{%
\pgfpathmoveto{\pgfqpoint{0.000000in}{0.000000in}}%
\pgfpathlineto{\pgfqpoint{0.000000in}{-0.048611in}}%
\pgfusepath{stroke,fill}%
}%
\begin{pgfscope}%
\pgfsys@transformshift{2.325513in}{0.431111in}%
\pgfsys@useobject{currentmarker}{}%
\end{pgfscope}%
\end{pgfscope}%
\begin{pgfscope}%
\definecolor{textcolor}{rgb}{0.000000,0.000000,0.000000}%
\pgfsetstrokecolor{textcolor}%
\pgfsetfillcolor{textcolor}%
\pgftext[x=2.325513in,y=0.333889in,,top]{\color{textcolor}\sffamily\fontsize{6.000000}{7.200000}\selectfont 19}%
\end{pgfscope}%
\begin{pgfscope}%
\pgfsetbuttcap%
\pgfsetroundjoin%
\definecolor{currentfill}{rgb}{0.000000,0.000000,0.000000}%
\pgfsetfillcolor{currentfill}%
\pgfsetlinewidth{0.401500pt}%
\definecolor{currentstroke}{rgb}{0.000000,0.000000,0.000000}%
\pgfsetstrokecolor{currentstroke}%
\pgfsetdash{}{0pt}%
\pgfsys@defobject{currentmarker}{\pgfqpoint{0.000000in}{-0.048611in}}{\pgfqpoint{0.000000in}{0.000000in}}{%
\pgfpathmoveto{\pgfqpoint{0.000000in}{0.000000in}}%
\pgfpathlineto{\pgfqpoint{0.000000in}{-0.048611in}}%
\pgfusepath{stroke,fill}%
}%
\begin{pgfscope}%
\pgfsys@transformshift{2.555511in}{0.431111in}%
\pgfsys@useobject{currentmarker}{}%
\end{pgfscope}%
\end{pgfscope}%
\begin{pgfscope}%
\definecolor{textcolor}{rgb}{0.000000,0.000000,0.000000}%
\pgfsetstrokecolor{textcolor}%
\pgfsetfillcolor{textcolor}%
\pgftext[x=2.555511in,y=0.333889in,,top]{\color{textcolor}\sffamily\fontsize{6.000000}{7.200000}\selectfont 21}%
\end{pgfscope}%
\begin{pgfscope}%
\pgfsetbuttcap%
\pgfsetroundjoin%
\definecolor{currentfill}{rgb}{0.000000,0.000000,0.000000}%
\pgfsetfillcolor{currentfill}%
\pgfsetlinewidth{0.401500pt}%
\definecolor{currentstroke}{rgb}{0.000000,0.000000,0.000000}%
\pgfsetstrokecolor{currentstroke}%
\pgfsetdash{}{0pt}%
\pgfsys@defobject{currentmarker}{\pgfqpoint{0.000000in}{-0.048611in}}{\pgfqpoint{0.000000in}{0.000000in}}{%
\pgfpathmoveto{\pgfqpoint{0.000000in}{0.000000in}}%
\pgfpathlineto{\pgfqpoint{0.000000in}{-0.048611in}}%
\pgfusepath{stroke,fill}%
}%
\begin{pgfscope}%
\pgfsys@transformshift{2.785509in}{0.431111in}%
\pgfsys@useobject{currentmarker}{}%
\end{pgfscope}%
\end{pgfscope}%
\begin{pgfscope}%
\definecolor{textcolor}{rgb}{0.000000,0.000000,0.000000}%
\pgfsetstrokecolor{textcolor}%
\pgfsetfillcolor{textcolor}%
\pgftext[x=2.785509in,y=0.333889in,,top]{\color{textcolor}\sffamily\fontsize{6.000000}{7.200000}\selectfont 23}%
\end{pgfscope}%
\begin{pgfscope}%
\pgfsetbuttcap%
\pgfsetroundjoin%
\definecolor{currentfill}{rgb}{0.000000,0.000000,0.000000}%
\pgfsetfillcolor{currentfill}%
\pgfsetlinewidth{0.401500pt}%
\definecolor{currentstroke}{rgb}{0.000000,0.000000,0.000000}%
\pgfsetstrokecolor{currentstroke}%
\pgfsetdash{}{0pt}%
\pgfsys@defobject{currentmarker}{\pgfqpoint{0.000000in}{-0.048611in}}{\pgfqpoint{0.000000in}{0.000000in}}{%
\pgfpathmoveto{\pgfqpoint{0.000000in}{0.000000in}}%
\pgfpathlineto{\pgfqpoint{0.000000in}{-0.048611in}}%
\pgfusepath{stroke,fill}%
}%
\begin{pgfscope}%
\pgfsys@transformshift{3.015507in}{0.431111in}%
\pgfsys@useobject{currentmarker}{}%
\end{pgfscope}%
\end{pgfscope}%
\begin{pgfscope}%
\definecolor{textcolor}{rgb}{0.000000,0.000000,0.000000}%
\pgfsetstrokecolor{textcolor}%
\pgfsetfillcolor{textcolor}%
\pgftext[x=3.015507in,y=0.333889in,,top]{\color{textcolor}\sffamily\fontsize{6.000000}{7.200000}\selectfont 25}%
\end{pgfscope}%
\begin{pgfscope}%
\pgfsetbuttcap%
\pgfsetroundjoin%
\definecolor{currentfill}{rgb}{0.000000,0.000000,0.000000}%
\pgfsetfillcolor{currentfill}%
\pgfsetlinewidth{0.401500pt}%
\definecolor{currentstroke}{rgb}{0.000000,0.000000,0.000000}%
\pgfsetstrokecolor{currentstroke}%
\pgfsetdash{}{0pt}%
\pgfsys@defobject{currentmarker}{\pgfqpoint{0.000000in}{-0.048611in}}{\pgfqpoint{0.000000in}{0.000000in}}{%
\pgfpathmoveto{\pgfqpoint{0.000000in}{0.000000in}}%
\pgfpathlineto{\pgfqpoint{0.000000in}{-0.048611in}}%
\pgfusepath{stroke,fill}%
}%
\begin{pgfscope}%
\pgfsys@transformshift{3.245505in}{0.431111in}%
\pgfsys@useobject{currentmarker}{}%
\end{pgfscope}%
\end{pgfscope}%
\begin{pgfscope}%
\definecolor{textcolor}{rgb}{0.000000,0.000000,0.000000}%
\pgfsetstrokecolor{textcolor}%
\pgfsetfillcolor{textcolor}%
\pgftext[x=3.245505in,y=0.333889in,,top]{\color{textcolor}\sffamily\fontsize{6.000000}{7.200000}\selectfont 27}%
\end{pgfscope}%
\begin{pgfscope}%
\definecolor{textcolor}{rgb}{0.000000,0.000000,0.000000}%
\pgfsetstrokecolor{textcolor}%
\pgfsetfillcolor{textcolor}%
\pgftext[x=1.980516in,y=0.204259in,,top]{\color{textcolor}\sffamily\fontsize{7.000000}{8.400000}\selectfont Number of Training Subjects}%
\end{pgfscope}%
\begin{pgfscope}%
\pgfpathrectangle{\pgfqpoint{0.589028in}{0.431111in}}{\pgfqpoint{2.782976in}{1.307222in}}%
\pgfusepath{clip}%
\pgfsetbuttcap%
\pgfsetroundjoin%
\pgfsetlinewidth{0.401500pt}%
\definecolor{currentstroke}{rgb}{0.647059,0.647059,0.647059}%
\pgfsetstrokecolor{currentstroke}%
\pgfsetdash{}{0pt}%
\pgfpathmoveto{\pgfqpoint{0.589028in}{0.431111in}}%
\pgfpathlineto{\pgfqpoint{3.372004in}{0.431111in}}%
\pgfusepath{stroke}%
\end{pgfscope}%
\begin{pgfscope}%
\pgfsetbuttcap%
\pgfsetroundjoin%
\definecolor{currentfill}{rgb}{0.000000,0.000000,0.000000}%
\pgfsetfillcolor{currentfill}%
\pgfsetlinewidth{0.401500pt}%
\definecolor{currentstroke}{rgb}{0.000000,0.000000,0.000000}%
\pgfsetstrokecolor{currentstroke}%
\pgfsetdash{}{0pt}%
\pgfsys@defobject{currentmarker}{\pgfqpoint{-0.048611in}{0.000000in}}{\pgfqpoint{-0.000000in}{0.000000in}}{%
\pgfpathmoveto{\pgfqpoint{-0.000000in}{0.000000in}}%
\pgfpathlineto{\pgfqpoint{-0.048611in}{0.000000in}}%
\pgfusepath{stroke,fill}%
}%
\begin{pgfscope}%
\pgfsys@transformshift{0.589028in}{0.431111in}%
\pgfsys@useobject{currentmarker}{}%
\end{pgfscope}%
\end{pgfscope}%
\begin{pgfscope}%
\definecolor{textcolor}{rgb}{0.000000,0.000000,0.000000}%
\pgfsetstrokecolor{textcolor}%
\pgfsetfillcolor{textcolor}%
\pgftext[x=0.373748in, y=0.402176in, left, base]{\color{textcolor}\sffamily\fontsize{6.000000}{7.200000}\selectfont 0\%}%
\end{pgfscope}%
\begin{pgfscope}%
\pgfpathrectangle{\pgfqpoint{0.589028in}{0.431111in}}{\pgfqpoint{2.782976in}{1.307222in}}%
\pgfusepath{clip}%
\pgfsetbuttcap%
\pgfsetroundjoin%
\pgfsetlinewidth{0.401500pt}%
\definecolor{currentstroke}{rgb}{0.647059,0.647059,0.647059}%
\pgfsetstrokecolor{currentstroke}%
\pgfsetdash{}{0pt}%
\pgfpathmoveto{\pgfqpoint{0.589028in}{0.692556in}}%
\pgfpathlineto{\pgfqpoint{3.372004in}{0.692556in}}%
\pgfusepath{stroke}%
\end{pgfscope}%
\begin{pgfscope}%
\pgfsetbuttcap%
\pgfsetroundjoin%
\definecolor{currentfill}{rgb}{0.000000,0.000000,0.000000}%
\pgfsetfillcolor{currentfill}%
\pgfsetlinewidth{0.401500pt}%
\definecolor{currentstroke}{rgb}{0.000000,0.000000,0.000000}%
\pgfsetstrokecolor{currentstroke}%
\pgfsetdash{}{0pt}%
\pgfsys@defobject{currentmarker}{\pgfqpoint{-0.048611in}{0.000000in}}{\pgfqpoint{-0.000000in}{0.000000in}}{%
\pgfpathmoveto{\pgfqpoint{-0.000000in}{0.000000in}}%
\pgfpathlineto{\pgfqpoint{-0.048611in}{0.000000in}}%
\pgfusepath{stroke,fill}%
}%
\begin{pgfscope}%
\pgfsys@transformshift{0.589028in}{0.692556in}%
\pgfsys@useobject{currentmarker}{}%
\end{pgfscope}%
\end{pgfscope}%
\begin{pgfscope}%
\definecolor{textcolor}{rgb}{0.000000,0.000000,0.000000}%
\pgfsetstrokecolor{textcolor}%
\pgfsetfillcolor{textcolor}%
\pgftext[x=0.329477in, y=0.663620in, left, base]{\color{textcolor}\sffamily\fontsize{6.000000}{7.200000}\selectfont 20\%}%
\end{pgfscope}%
\begin{pgfscope}%
\pgfpathrectangle{\pgfqpoint{0.589028in}{0.431111in}}{\pgfqpoint{2.782976in}{1.307222in}}%
\pgfusepath{clip}%
\pgfsetbuttcap%
\pgfsetroundjoin%
\pgfsetlinewidth{0.401500pt}%
\definecolor{currentstroke}{rgb}{0.647059,0.647059,0.647059}%
\pgfsetstrokecolor{currentstroke}%
\pgfsetdash{}{0pt}%
\pgfpathmoveto{\pgfqpoint{0.589028in}{0.954000in}}%
\pgfpathlineto{\pgfqpoint{3.372004in}{0.954000in}}%
\pgfusepath{stroke}%
\end{pgfscope}%
\begin{pgfscope}%
\pgfsetbuttcap%
\pgfsetroundjoin%
\definecolor{currentfill}{rgb}{0.000000,0.000000,0.000000}%
\pgfsetfillcolor{currentfill}%
\pgfsetlinewidth{0.401500pt}%
\definecolor{currentstroke}{rgb}{0.000000,0.000000,0.000000}%
\pgfsetstrokecolor{currentstroke}%
\pgfsetdash{}{0pt}%
\pgfsys@defobject{currentmarker}{\pgfqpoint{-0.048611in}{0.000000in}}{\pgfqpoint{-0.000000in}{0.000000in}}{%
\pgfpathmoveto{\pgfqpoint{-0.000000in}{0.000000in}}%
\pgfpathlineto{\pgfqpoint{-0.048611in}{0.000000in}}%
\pgfusepath{stroke,fill}%
}%
\begin{pgfscope}%
\pgfsys@transformshift{0.589028in}{0.954000in}%
\pgfsys@useobject{currentmarker}{}%
\end{pgfscope}%
\end{pgfscope}%
\begin{pgfscope}%
\definecolor{textcolor}{rgb}{0.000000,0.000000,0.000000}%
\pgfsetstrokecolor{textcolor}%
\pgfsetfillcolor{textcolor}%
\pgftext[x=0.329477in, y=0.925065in, left, base]{\color{textcolor}\sffamily\fontsize{6.000000}{7.200000}\selectfont 40\%}%
\end{pgfscope}%
\begin{pgfscope}%
\pgfpathrectangle{\pgfqpoint{0.589028in}{0.431111in}}{\pgfqpoint{2.782976in}{1.307222in}}%
\pgfusepath{clip}%
\pgfsetbuttcap%
\pgfsetroundjoin%
\pgfsetlinewidth{0.401500pt}%
\definecolor{currentstroke}{rgb}{0.647059,0.647059,0.647059}%
\pgfsetstrokecolor{currentstroke}%
\pgfsetdash{}{0pt}%
\pgfpathmoveto{\pgfqpoint{0.589028in}{1.215444in}}%
\pgfpathlineto{\pgfqpoint{3.372004in}{1.215444in}}%
\pgfusepath{stroke}%
\end{pgfscope}%
\begin{pgfscope}%
\pgfsetbuttcap%
\pgfsetroundjoin%
\definecolor{currentfill}{rgb}{0.000000,0.000000,0.000000}%
\pgfsetfillcolor{currentfill}%
\pgfsetlinewidth{0.401500pt}%
\definecolor{currentstroke}{rgb}{0.000000,0.000000,0.000000}%
\pgfsetstrokecolor{currentstroke}%
\pgfsetdash{}{0pt}%
\pgfsys@defobject{currentmarker}{\pgfqpoint{-0.048611in}{0.000000in}}{\pgfqpoint{-0.000000in}{0.000000in}}{%
\pgfpathmoveto{\pgfqpoint{-0.000000in}{0.000000in}}%
\pgfpathlineto{\pgfqpoint{-0.048611in}{0.000000in}}%
\pgfusepath{stroke,fill}%
}%
\begin{pgfscope}%
\pgfsys@transformshift{0.589028in}{1.215444in}%
\pgfsys@useobject{currentmarker}{}%
\end{pgfscope}%
\end{pgfscope}%
\begin{pgfscope}%
\definecolor{textcolor}{rgb}{0.000000,0.000000,0.000000}%
\pgfsetstrokecolor{textcolor}%
\pgfsetfillcolor{textcolor}%
\pgftext[x=0.329477in, y=1.186509in, left, base]{\color{textcolor}\sffamily\fontsize{6.000000}{7.200000}\selectfont 60\%}%
\end{pgfscope}%
\begin{pgfscope}%
\pgfpathrectangle{\pgfqpoint{0.589028in}{0.431111in}}{\pgfqpoint{2.782976in}{1.307222in}}%
\pgfusepath{clip}%
\pgfsetbuttcap%
\pgfsetroundjoin%
\pgfsetlinewidth{0.401500pt}%
\definecolor{currentstroke}{rgb}{0.647059,0.647059,0.647059}%
\pgfsetstrokecolor{currentstroke}%
\pgfsetdash{}{0pt}%
\pgfpathmoveto{\pgfqpoint{0.589028in}{1.476889in}}%
\pgfpathlineto{\pgfqpoint{3.372004in}{1.476889in}}%
\pgfusepath{stroke}%
\end{pgfscope}%
\begin{pgfscope}%
\pgfsetbuttcap%
\pgfsetroundjoin%
\definecolor{currentfill}{rgb}{0.000000,0.000000,0.000000}%
\pgfsetfillcolor{currentfill}%
\pgfsetlinewidth{0.401500pt}%
\definecolor{currentstroke}{rgb}{0.000000,0.000000,0.000000}%
\pgfsetstrokecolor{currentstroke}%
\pgfsetdash{}{0pt}%
\pgfsys@defobject{currentmarker}{\pgfqpoint{-0.048611in}{0.000000in}}{\pgfqpoint{-0.000000in}{0.000000in}}{%
\pgfpathmoveto{\pgfqpoint{-0.000000in}{0.000000in}}%
\pgfpathlineto{\pgfqpoint{-0.048611in}{0.000000in}}%
\pgfusepath{stroke,fill}%
}%
\begin{pgfscope}%
\pgfsys@transformshift{0.589028in}{1.476889in}%
\pgfsys@useobject{currentmarker}{}%
\end{pgfscope}%
\end{pgfscope}%
\begin{pgfscope}%
\definecolor{textcolor}{rgb}{0.000000,0.000000,0.000000}%
\pgfsetstrokecolor{textcolor}%
\pgfsetfillcolor{textcolor}%
\pgftext[x=0.329477in, y=1.447954in, left, base]{\color{textcolor}\sffamily\fontsize{6.000000}{7.200000}\selectfont 80\%}%
\end{pgfscope}%
\begin{pgfscope}%
\pgfpathrectangle{\pgfqpoint{0.589028in}{0.431111in}}{\pgfqpoint{2.782976in}{1.307222in}}%
\pgfusepath{clip}%
\pgfsetbuttcap%
\pgfsetroundjoin%
\pgfsetlinewidth{0.401500pt}%
\definecolor{currentstroke}{rgb}{0.647059,0.647059,0.647059}%
\pgfsetstrokecolor{currentstroke}%
\pgfsetdash{}{0pt}%
\pgfpathmoveto{\pgfqpoint{0.589028in}{1.738333in}}%
\pgfpathlineto{\pgfqpoint{3.372004in}{1.738333in}}%
\pgfusepath{stroke}%
\end{pgfscope}%
\begin{pgfscope}%
\pgfsetbuttcap%
\pgfsetroundjoin%
\definecolor{currentfill}{rgb}{0.000000,0.000000,0.000000}%
\pgfsetfillcolor{currentfill}%
\pgfsetlinewidth{0.401500pt}%
\definecolor{currentstroke}{rgb}{0.000000,0.000000,0.000000}%
\pgfsetstrokecolor{currentstroke}%
\pgfsetdash{}{0pt}%
\pgfsys@defobject{currentmarker}{\pgfqpoint{-0.048611in}{0.000000in}}{\pgfqpoint{-0.000000in}{0.000000in}}{%
\pgfpathmoveto{\pgfqpoint{-0.000000in}{0.000000in}}%
\pgfpathlineto{\pgfqpoint{-0.048611in}{0.000000in}}%
\pgfusepath{stroke,fill}%
}%
\begin{pgfscope}%
\pgfsys@transformshift{0.589028in}{1.738333in}%
\pgfsys@useobject{currentmarker}{}%
\end{pgfscope}%
\end{pgfscope}%
\begin{pgfscope}%
\definecolor{textcolor}{rgb}{0.000000,0.000000,0.000000}%
\pgfsetstrokecolor{textcolor}%
\pgfsetfillcolor{textcolor}%
\pgftext[x=0.285205in, y=1.709398in, left, base]{\color{textcolor}\sffamily\fontsize{6.000000}{7.200000}\selectfont 100\%}%
\end{pgfscope}%
\begin{pgfscope}%
\definecolor{textcolor}{rgb}{0.000000,0.000000,0.000000}%
\pgfsetstrokecolor{textcolor}%
\pgfsetfillcolor{textcolor}%
\pgftext[x=0.229650in,y=1.084722in,,bottom,rotate=90.000000]{\color{textcolor}\sffamily\fontsize{7.000000}{8.400000}\selectfont Accuracy}%
\end{pgfscope}%
\begin{pgfscope}%
\pgfpathrectangle{\pgfqpoint{0.589028in}{0.431111in}}{\pgfqpoint{2.782976in}{1.307222in}}%
\pgfusepath{clip}%
\pgfsetbuttcap%
\pgfsetroundjoin%
\definecolor{currentfill}{rgb}{1.000000,0.650980,0.000000}%
\pgfsetfillcolor{currentfill}%
\pgfsetfillopacity{0.800000}%
\pgfsetlinewidth{1.003750pt}%
\definecolor{currentstroke}{rgb}{1.000000,0.650980,0.000000}%
\pgfsetstrokecolor{currentstroke}%
\pgfsetstrokeopacity{0.800000}%
\pgfsetdash{}{0pt}%
\pgfsys@defobject{currentmarker}{\pgfqpoint{-0.041667in}{-0.041667in}}{\pgfqpoint{0.041667in}{0.041667in}}{%
\pgfpathmoveto{\pgfqpoint{0.000000in}{-0.041667in}}%
\pgfpathcurveto{\pgfqpoint{0.011050in}{-0.041667in}}{\pgfqpoint{0.021649in}{-0.037276in}}{\pgfqpoint{0.029463in}{-0.029463in}}%
\pgfpathcurveto{\pgfqpoint{0.037276in}{-0.021649in}}{\pgfqpoint{0.041667in}{-0.011050in}}{\pgfqpoint{0.041667in}{0.000000in}}%
\pgfpathcurveto{\pgfqpoint{0.041667in}{0.011050in}}{\pgfqpoint{0.037276in}{0.021649in}}{\pgfqpoint{0.029463in}{0.029463in}}%
\pgfpathcurveto{\pgfqpoint{0.021649in}{0.037276in}}{\pgfqpoint{0.011050in}{0.041667in}}{\pgfqpoint{0.000000in}{0.041667in}}%
\pgfpathcurveto{\pgfqpoint{-0.011050in}{0.041667in}}{\pgfqpoint{-0.021649in}{0.037276in}}{\pgfqpoint{-0.029463in}{0.029463in}}%
\pgfpathcurveto{\pgfqpoint{-0.037276in}{0.021649in}}{\pgfqpoint{-0.041667in}{0.011050in}}{\pgfqpoint{-0.041667in}{0.000000in}}%
\pgfpathcurveto{\pgfqpoint{-0.041667in}{-0.011050in}}{\pgfqpoint{-0.037276in}{-0.021649in}}{\pgfqpoint{-0.029463in}{-0.029463in}}%
\pgfpathcurveto{\pgfqpoint{-0.021649in}{-0.037276in}}{\pgfqpoint{-0.011050in}{-0.041667in}}{\pgfqpoint{0.000000in}{-0.041667in}}%
\pgfpathlineto{\pgfqpoint{0.000000in}{-0.041667in}}%
\pgfpathclose%
\pgfusepath{stroke,fill}%
}%
\begin{pgfscope}%
\pgfsys@transformshift{3.245505in}{1.516079in}%
\pgfsys@useobject{currentmarker}{}%
\end{pgfscope}%
\begin{pgfscope}%
\pgfsys@transformshift{1.175523in}{1.172528in}%
\pgfsys@useobject{currentmarker}{}%
\end{pgfscope}%
\begin{pgfscope}%
\pgfsys@transformshift{0.945525in}{0.907360in}%
\pgfsys@useobject{currentmarker}{}%
\end{pgfscope}%
\begin{pgfscope}%
\pgfsys@transformshift{0.715527in}{0.787865in}%
\pgfsys@useobject{currentmarker}{}%
\end{pgfscope}%
\begin{pgfscope}%
\pgfsys@transformshift{3.015507in}{1.485665in}%
\pgfsys@useobject{currentmarker}{}%
\end{pgfscope}%
\begin{pgfscope}%
\pgfsys@transformshift{2.785509in}{1.398145in}%
\pgfsys@useobject{currentmarker}{}%
\end{pgfscope}%
\begin{pgfscope}%
\pgfsys@transformshift{2.555511in}{1.375248in}%
\pgfsys@useobject{currentmarker}{}%
\end{pgfscope}%
\begin{pgfscope}%
\pgfsys@transformshift{2.325513in}{1.377100in}%
\pgfsys@useobject{currentmarker}{}%
\end{pgfscope}%
\begin{pgfscope}%
\pgfsys@transformshift{2.095515in}{1.221530in}%
\pgfsys@useobject{currentmarker}{}%
\end{pgfscope}%
\begin{pgfscope}%
\pgfsys@transformshift{1.865517in}{1.276576in}%
\pgfsys@useobject{currentmarker}{}%
\end{pgfscope}%
\begin{pgfscope}%
\pgfsys@transformshift{1.635519in}{1.345248in}%
\pgfsys@useobject{currentmarker}{}%
\end{pgfscope}%
\begin{pgfscope}%
\pgfsys@transformshift{1.405521in}{1.148968in}%
\pgfsys@useobject{currentmarker}{}%
\end{pgfscope}%
\end{pgfscope}%
\begin{pgfscope}%
\pgfpathrectangle{\pgfqpoint{0.589028in}{0.431111in}}{\pgfqpoint{2.782976in}{1.307222in}}%
\pgfusepath{clip}%
\pgfsetbuttcap%
\pgfsetroundjoin%
\pgfsetlinewidth{1.505625pt}%
\definecolor{currentstroke}{rgb}{1.000000,0.650980,0.000000}%
\pgfsetstrokecolor{currentstroke}%
\pgfsetdash{{1.500000pt}{2.475000pt}}{0.000000pt}%
\pgfpathmoveto{\pgfqpoint{0.715527in}{0.943981in}}%
\pgfpathlineto{\pgfqpoint{0.741082in}{0.950184in}}%
\pgfpathlineto{\pgfqpoint{0.766637in}{0.956387in}}%
\pgfpathlineto{\pgfqpoint{0.792193in}{0.962590in}}%
\pgfpathlineto{\pgfqpoint{0.817748in}{0.968793in}}%
\pgfpathlineto{\pgfqpoint{0.843303in}{0.974996in}}%
\pgfpathlineto{\pgfqpoint{0.868859in}{0.981199in}}%
\pgfpathlineto{\pgfqpoint{0.894414in}{0.987402in}}%
\pgfpathlineto{\pgfqpoint{0.919969in}{0.993605in}}%
\pgfpathlineto{\pgfqpoint{0.945525in}{0.999807in}}%
\pgfpathlineto{\pgfqpoint{0.971080in}{1.006010in}}%
\pgfpathlineto{\pgfqpoint{0.996635in}{1.012213in}}%
\pgfpathlineto{\pgfqpoint{1.022191in}{1.018416in}}%
\pgfpathlineto{\pgfqpoint{1.047746in}{1.024619in}}%
\pgfpathlineto{\pgfqpoint{1.073301in}{1.030822in}}%
\pgfpathlineto{\pgfqpoint{1.098857in}{1.037025in}}%
\pgfpathlineto{\pgfqpoint{1.124412in}{1.043228in}}%
\pgfpathlineto{\pgfqpoint{1.149967in}{1.049431in}}%
\pgfpathlineto{\pgfqpoint{1.175523in}{1.055634in}}%
\pgfpathlineto{\pgfqpoint{1.201078in}{1.061837in}}%
\pgfpathlineto{\pgfqpoint{1.226633in}{1.068040in}}%
\pgfpathlineto{\pgfqpoint{1.252189in}{1.074243in}}%
\pgfpathlineto{\pgfqpoint{1.277744in}{1.080445in}}%
\pgfpathlineto{\pgfqpoint{1.303299in}{1.086648in}}%
\pgfpathlineto{\pgfqpoint{1.328855in}{1.092851in}}%
\pgfpathlineto{\pgfqpoint{1.354410in}{1.099054in}}%
\pgfpathlineto{\pgfqpoint{1.379965in}{1.105257in}}%
\pgfpathlineto{\pgfqpoint{1.405521in}{1.111460in}}%
\pgfpathlineto{\pgfqpoint{1.431076in}{1.117663in}}%
\pgfpathlineto{\pgfqpoint{1.456631in}{1.123866in}}%
\pgfpathlineto{\pgfqpoint{1.482187in}{1.130069in}}%
\pgfpathlineto{\pgfqpoint{1.507742in}{1.136272in}}%
\pgfpathlineto{\pgfqpoint{1.533297in}{1.142475in}}%
\pgfpathlineto{\pgfqpoint{1.558853in}{1.148678in}}%
\pgfpathlineto{\pgfqpoint{1.584408in}{1.154881in}}%
\pgfpathlineto{\pgfqpoint{1.609963in}{1.161083in}}%
\pgfpathlineto{\pgfqpoint{1.635519in}{1.167286in}}%
\pgfpathlineto{\pgfqpoint{1.661074in}{1.173489in}}%
\pgfpathlineto{\pgfqpoint{1.686629in}{1.179692in}}%
\pgfpathlineto{\pgfqpoint{1.712185in}{1.185895in}}%
\pgfpathlineto{\pgfqpoint{1.737740in}{1.192098in}}%
\pgfpathlineto{\pgfqpoint{1.763295in}{1.198301in}}%
\pgfpathlineto{\pgfqpoint{1.788851in}{1.204504in}}%
\pgfpathlineto{\pgfqpoint{1.814406in}{1.210707in}}%
\pgfpathlineto{\pgfqpoint{1.839962in}{1.216910in}}%
\pgfpathlineto{\pgfqpoint{1.865517in}{1.223113in}}%
\pgfpathlineto{\pgfqpoint{1.891072in}{1.229316in}}%
\pgfpathlineto{\pgfqpoint{1.916628in}{1.235519in}}%
\pgfpathlineto{\pgfqpoint{1.942183in}{1.241721in}}%
\pgfpathlineto{\pgfqpoint{1.967738in}{1.247924in}}%
\pgfpathlineto{\pgfqpoint{1.993294in}{1.254127in}}%
\pgfpathlineto{\pgfqpoint{2.018849in}{1.260330in}}%
\pgfpathlineto{\pgfqpoint{2.044404in}{1.266533in}}%
\pgfpathlineto{\pgfqpoint{2.069960in}{1.272736in}}%
\pgfpathlineto{\pgfqpoint{2.095515in}{1.278939in}}%
\pgfpathlineto{\pgfqpoint{2.121070in}{1.285142in}}%
\pgfpathlineto{\pgfqpoint{2.146626in}{1.291345in}}%
\pgfpathlineto{\pgfqpoint{2.172181in}{1.297548in}}%
\pgfpathlineto{\pgfqpoint{2.197736in}{1.303751in}}%
\pgfpathlineto{\pgfqpoint{2.223292in}{1.309954in}}%
\pgfpathlineto{\pgfqpoint{2.248847in}{1.316157in}}%
\pgfpathlineto{\pgfqpoint{2.274402in}{1.322359in}}%
\pgfpathlineto{\pgfqpoint{2.299958in}{1.328562in}}%
\pgfpathlineto{\pgfqpoint{2.325513in}{1.334765in}}%
\pgfpathlineto{\pgfqpoint{2.351068in}{1.340968in}}%
\pgfpathlineto{\pgfqpoint{2.376624in}{1.347171in}}%
\pgfpathlineto{\pgfqpoint{2.402179in}{1.353374in}}%
\pgfpathlineto{\pgfqpoint{2.427734in}{1.359577in}}%
\pgfpathlineto{\pgfqpoint{2.453290in}{1.365780in}}%
\pgfpathlineto{\pgfqpoint{2.478845in}{1.371983in}}%
\pgfpathlineto{\pgfqpoint{2.504400in}{1.378186in}}%
\pgfpathlineto{\pgfqpoint{2.529956in}{1.384389in}}%
\pgfpathlineto{\pgfqpoint{2.555511in}{1.390592in}}%
\pgfpathlineto{\pgfqpoint{2.581066in}{1.396795in}}%
\pgfpathlineto{\pgfqpoint{2.606622in}{1.402997in}}%
\pgfpathlineto{\pgfqpoint{2.632177in}{1.409200in}}%
\pgfpathlineto{\pgfqpoint{2.657732in}{1.415403in}}%
\pgfpathlineto{\pgfqpoint{2.683288in}{1.421606in}}%
\pgfpathlineto{\pgfqpoint{2.708843in}{1.427809in}}%
\pgfpathlineto{\pgfqpoint{2.734398in}{1.434012in}}%
\pgfpathlineto{\pgfqpoint{2.759954in}{1.440215in}}%
\pgfpathlineto{\pgfqpoint{2.785509in}{1.446418in}}%
\pgfpathlineto{\pgfqpoint{2.811064in}{1.452621in}}%
\pgfpathlineto{\pgfqpoint{2.836620in}{1.458824in}}%
\pgfpathlineto{\pgfqpoint{2.862175in}{1.465027in}}%
\pgfpathlineto{\pgfqpoint{2.887730in}{1.471230in}}%
\pgfpathlineto{\pgfqpoint{2.913286in}{1.477433in}}%
\pgfpathlineto{\pgfqpoint{2.938841in}{1.483636in}}%
\pgfpathlineto{\pgfqpoint{2.964396in}{1.489838in}}%
\pgfpathlineto{\pgfqpoint{2.989952in}{1.496041in}}%
\pgfpathlineto{\pgfqpoint{3.015507in}{1.502244in}}%
\pgfpathlineto{\pgfqpoint{3.041062in}{1.508447in}}%
\pgfpathlineto{\pgfqpoint{3.066618in}{1.514650in}}%
\pgfpathlineto{\pgfqpoint{3.092173in}{1.520853in}}%
\pgfpathlineto{\pgfqpoint{3.117728in}{1.527056in}}%
\pgfpathlineto{\pgfqpoint{3.143284in}{1.533259in}}%
\pgfpathlineto{\pgfqpoint{3.168839in}{1.539462in}}%
\pgfpathlineto{\pgfqpoint{3.194394in}{1.545665in}}%
\pgfpathlineto{\pgfqpoint{3.219950in}{1.551868in}}%
\pgfpathlineto{\pgfqpoint{3.245505in}{1.558071in}}%
\pgfusepath{stroke}%
\end{pgfscope}%
\begin{pgfscope}%
\pgfsetrectcap%
\pgfsetmiterjoin%
\pgfsetlinewidth{1.003750pt}%
\definecolor{currentstroke}{rgb}{0.000000,0.000000,0.000000}%
\pgfsetstrokecolor{currentstroke}%
\pgfsetdash{}{0pt}%
\pgfpathmoveto{\pgfqpoint{0.589028in}{0.431111in}}%
\pgfpathlineto{\pgfqpoint{0.589028in}{1.738333in}}%
\pgfusepath{stroke}%
\end{pgfscope}%
\begin{pgfscope}%
\pgfsetrectcap%
\pgfsetmiterjoin%
\pgfsetlinewidth{1.003750pt}%
\definecolor{currentstroke}{rgb}{0.000000,0.000000,0.000000}%
\pgfsetstrokecolor{currentstroke}%
\pgfsetdash{}{0pt}%
\pgfpathmoveto{\pgfqpoint{0.589028in}{0.431111in}}%
\pgfpathlineto{\pgfqpoint{3.372004in}{0.431111in}}%
\pgfusepath{stroke}%
\end{pgfscope}%
\begin{pgfscope}%
\definecolor{textcolor}{rgb}{0.000000,0.000000,0.000000}%
\pgfsetstrokecolor{textcolor}%
\pgfsetfillcolor{textcolor}%
\pgftext[x=1.980516in,y=1.821667in,,base]{\color{textcolor}\sffamily\fontsize{7.000000}{8.400000}\selectfont More Users Improve Identification Performance}%
\end{pgfscope}%
\end{pgfpicture}%
\makeatother%
\endgroup%

%% file: results/generated/miller_evaluation.pgf
\begingroup%
\makeatletter%
\begin{pgfpicture}%
\pgfpathrectangle{\pgfpointorigin}{\pgfqpoint{3.477004in}{2.500000in}}%
\pgfusepath{use as bounding box, clip}%
\begin{pgfscope}%
\pgfsetbuttcap%
\pgfsetmiterjoin%
\definecolor{currentfill}{rgb}{1.000000,1.000000,1.000000}%
\pgfsetfillcolor{currentfill}%
\pgfsetlinewidth{0.000000pt}%
\definecolor{currentstroke}{rgb}{1.000000,1.000000,1.000000}%
\pgfsetstrokecolor{currentstroke}%
\pgfsetdash{}{0pt}%
\pgfpathmoveto{\pgfqpoint{0.000000in}{0.000000in}}%
\pgfpathlineto{\pgfqpoint{3.477004in}{0.000000in}}%
\pgfpathlineto{\pgfqpoint{3.477004in}{2.500000in}}%
\pgfpathlineto{\pgfqpoint{0.000000in}{2.500000in}}%
\pgfpathlineto{\pgfqpoint{0.000000in}{0.000000in}}%
\pgfpathclose%
\pgfusepath{fill}%
\end{pgfscope}%
\begin{pgfscope}%
\pgfsetbuttcap%
\pgfsetmiterjoin%
\definecolor{currentfill}{rgb}{1.000000,1.000000,1.000000}%
\pgfsetfillcolor{currentfill}%
\pgfsetlinewidth{0.000000pt}%
\definecolor{currentstroke}{rgb}{0.000000,0.000000,0.000000}%
\pgfsetstrokecolor{currentstroke}%
\pgfsetstrokeopacity{0.000000}%
\pgfsetdash{}{0pt}%
\pgfpathmoveto{\pgfqpoint{0.536528in}{0.633889in}}%
\pgfpathlineto{\pgfqpoint{3.372004in}{0.633889in}}%
\pgfpathlineto{\pgfqpoint{3.372004in}{2.238333in}}%
\pgfpathlineto{\pgfqpoint{0.536528in}{2.238333in}}%
\pgfpathlineto{\pgfqpoint{0.536528in}{0.633889in}}%
\pgfpathclose%
\pgfusepath{fill}%
\end{pgfscope}%
\begin{pgfscope}%
\pgfsetbuttcap%
\pgfsetroundjoin%
\definecolor{currentfill}{rgb}{0.000000,0.000000,0.000000}%
\pgfsetfillcolor{currentfill}%
\pgfsetlinewidth{0.401500pt}%
\definecolor{currentstroke}{rgb}{0.000000,0.000000,0.000000}%
\pgfsetstrokecolor{currentstroke}%
\pgfsetdash{}{0pt}%
\pgfsys@defobject{currentmarker}{\pgfqpoint{0.000000in}{-0.048611in}}{\pgfqpoint{0.000000in}{0.000000in}}{%
\pgfpathmoveto{\pgfqpoint{0.000000in}{0.000000in}}%
\pgfpathlineto{\pgfqpoint{0.000000in}{-0.048611in}}%
\pgfusepath{stroke,fill}%
}%
\begin{pgfscope}%
\pgfsys@transformshift{0.941596in}{0.633889in}%
\pgfsys@useobject{currentmarker}{}%
\end{pgfscope}%
\end{pgfscope}%
\begin{pgfscope}%
\definecolor{textcolor}{rgb}{0.000000,0.000000,0.000000}%
\pgfsetstrokecolor{textcolor}%
\pgfsetfillcolor{textcolor}%
\pgftext[x=0.941596in,y=0.536667in,,top]{\color{textcolor}\sffamily\fontsize{6.000000}{7.200000}\selectfont Cosmos}%
\end{pgfscope}%
\begin{pgfscope}%
\pgfsetbuttcap%
\pgfsetroundjoin%
\definecolor{currentfill}{rgb}{0.000000,0.000000,0.000000}%
\pgfsetfillcolor{currentfill}%
\pgfsetlinewidth{0.401500pt}%
\definecolor{currentstroke}{rgb}{0.000000,0.000000,0.000000}%
\pgfsetstrokecolor{currentstroke}%
\pgfsetdash{}{0pt}%
\pgfsys@defobject{currentmarker}{\pgfqpoint{0.000000in}{-0.048611in}}{\pgfqpoint{0.000000in}{0.000000in}}{%
\pgfpathmoveto{\pgfqpoint{0.000000in}{0.000000in}}%
\pgfpathlineto{\pgfqpoint{0.000000in}{-0.048611in}}%
\pgfusepath{stroke,fill}%
}%
\begin{pgfscope}%
\pgfsys@transformshift{1.616709in}{0.633889in}%
\pgfsys@useobject{currentmarker}{}%
\end{pgfscope}%
\end{pgfscope}%
\begin{pgfscope}%
\definecolor{textcolor}{rgb}{0.000000,0.000000,0.000000}%
\pgfsetstrokecolor{textcolor}%
\pgfsetfillcolor{textcolor}%
\pgftext[x=1.616709in,y=0.536667in,,top]{\color{textcolor}\sffamily\fontsize{6.000000}{7.200000}\selectfont Quest}%
\end{pgfscope}%
\begin{pgfscope}%
\pgfsetbuttcap%
\pgfsetroundjoin%
\definecolor{currentfill}{rgb}{0.000000,0.000000,0.000000}%
\pgfsetfillcolor{currentfill}%
\pgfsetlinewidth{0.401500pt}%
\definecolor{currentstroke}{rgb}{0.000000,0.000000,0.000000}%
\pgfsetstrokecolor{currentstroke}%
\pgfsetdash{}{0pt}%
\pgfsys@defobject{currentmarker}{\pgfqpoint{0.000000in}{-0.048611in}}{\pgfqpoint{0.000000in}{0.000000in}}{%
\pgfpathmoveto{\pgfqpoint{0.000000in}{0.000000in}}%
\pgfpathlineto{\pgfqpoint{0.000000in}{-0.048611in}}%
\pgfusepath{stroke,fill}%
}%
\begin{pgfscope}%
\pgfsys@transformshift{2.291823in}{0.633889in}%
\pgfsys@useobject{currentmarker}{}%
\end{pgfscope}%
\end{pgfscope}%
\begin{pgfscope}%
\definecolor{textcolor}{rgb}{0.000000,0.000000,0.000000}%
\pgfsetstrokecolor{textcolor}%
\pgfsetfillcolor{textcolor}%
\pgftext[x=2.291823in,y=0.536667in,,top]{\color{textcolor}\sffamily\fontsize{6.000000}{7.200000}\selectfont Vive}%
\end{pgfscope}%
\begin{pgfscope}%
\pgfsetbuttcap%
\pgfsetroundjoin%
\definecolor{currentfill}{rgb}{0.000000,0.000000,0.000000}%
\pgfsetfillcolor{currentfill}%
\pgfsetlinewidth{0.401500pt}%
\definecolor{currentstroke}{rgb}{0.000000,0.000000,0.000000}%
\pgfsetstrokecolor{currentstroke}%
\pgfsetdash{}{0pt}%
\pgfsys@defobject{currentmarker}{\pgfqpoint{0.000000in}{-0.048611in}}{\pgfqpoint{0.000000in}{0.000000in}}{%
\pgfpathmoveto{\pgfqpoint{0.000000in}{0.000000in}}%
\pgfpathlineto{\pgfqpoint{0.000000in}{-0.048611in}}%
\pgfusepath{stroke,fill}%
}%
\begin{pgfscope}%
\pgfsys@transformshift{2.966936in}{0.633889in}%
\pgfsys@useobject{currentmarker}{}%
\end{pgfscope}%
\end{pgfscope}%
\begin{pgfscope}%
\definecolor{textcolor}{rgb}{0.000000,0.000000,0.000000}%
\pgfsetstrokecolor{textcolor}%
\pgfsetfillcolor{textcolor}%
\pgftext[x=2.966936in,y=0.536667in,,top]{\color{textcolor}\sffamily\fontsize{6.000000}{7.200000}\selectfont all}%
\end{pgfscope}%
\begin{pgfscope}%
\pgfpathrectangle{\pgfqpoint{0.536528in}{0.633889in}}{\pgfqpoint{2.835476in}{1.604444in}}%
\pgfusepath{clip}%
\pgfsetbuttcap%
\pgfsetroundjoin%
\pgfsetlinewidth{0.401500pt}%
\definecolor{currentstroke}{rgb}{0.647059,0.647059,0.647059}%
\pgfsetstrokecolor{currentstroke}%
\pgfsetdash{}{0pt}%
\pgfpathmoveto{\pgfqpoint{0.536528in}{0.832687in}}%
\pgfpathlineto{\pgfqpoint{3.372004in}{0.832687in}}%
\pgfusepath{stroke}%
\end{pgfscope}%
\begin{pgfscope}%
\pgfsetbuttcap%
\pgfsetroundjoin%
\definecolor{currentfill}{rgb}{0.000000,0.000000,0.000000}%
\pgfsetfillcolor{currentfill}%
\pgfsetlinewidth{0.401500pt}%
\definecolor{currentstroke}{rgb}{0.000000,0.000000,0.000000}%
\pgfsetstrokecolor{currentstroke}%
\pgfsetdash{}{0pt}%
\pgfsys@defobject{currentmarker}{\pgfqpoint{-0.048611in}{0.000000in}}{\pgfqpoint{-0.000000in}{0.000000in}}{%
\pgfpathmoveto{\pgfqpoint{-0.000000in}{0.000000in}}%
\pgfpathlineto{\pgfqpoint{-0.048611in}{0.000000in}}%
\pgfusepath{stroke,fill}%
}%
\begin{pgfscope}%
\pgfsys@transformshift{0.536528in}{0.832687in}%
\pgfsys@useobject{currentmarker}{}%
\end{pgfscope}%
\end{pgfscope}%
\begin{pgfscope}%
\definecolor{textcolor}{rgb}{0.000000,0.000000,0.000000}%
\pgfsetstrokecolor{textcolor}%
\pgfsetfillcolor{textcolor}%
\pgftext[x=0.276977in, y=0.803752in, left, base]{\color{textcolor}\sffamily\fontsize{6.000000}{7.200000}\selectfont 10\%}%
\end{pgfscope}%
\begin{pgfscope}%
\pgfpathrectangle{\pgfqpoint{0.536528in}{0.633889in}}{\pgfqpoint{2.835476in}{1.604444in}}%
\pgfusepath{clip}%
\pgfsetbuttcap%
\pgfsetroundjoin%
\pgfsetlinewidth{0.401500pt}%
\definecolor{currentstroke}{rgb}{0.647059,0.647059,0.647059}%
\pgfsetstrokecolor{currentstroke}%
\pgfsetdash{}{0pt}%
\pgfpathmoveto{\pgfqpoint{0.536528in}{1.033494in}}%
\pgfpathlineto{\pgfqpoint{3.372004in}{1.033494in}}%
\pgfusepath{stroke}%
\end{pgfscope}%
\begin{pgfscope}%
\pgfsetbuttcap%
\pgfsetroundjoin%
\definecolor{currentfill}{rgb}{0.000000,0.000000,0.000000}%
\pgfsetfillcolor{currentfill}%
\pgfsetlinewidth{0.401500pt}%
\definecolor{currentstroke}{rgb}{0.000000,0.000000,0.000000}%
\pgfsetstrokecolor{currentstroke}%
\pgfsetdash{}{0pt}%
\pgfsys@defobject{currentmarker}{\pgfqpoint{-0.048611in}{0.000000in}}{\pgfqpoint{-0.000000in}{0.000000in}}{%
\pgfpathmoveto{\pgfqpoint{-0.000000in}{0.000000in}}%
\pgfpathlineto{\pgfqpoint{-0.048611in}{0.000000in}}%
\pgfusepath{stroke,fill}%
}%
\begin{pgfscope}%
\pgfsys@transformshift{0.536528in}{1.033494in}%
\pgfsys@useobject{currentmarker}{}%
\end{pgfscope}%
\end{pgfscope}%
\begin{pgfscope}%
\definecolor{textcolor}{rgb}{0.000000,0.000000,0.000000}%
\pgfsetstrokecolor{textcolor}%
\pgfsetfillcolor{textcolor}%
\pgftext[x=0.276977in, y=1.004559in, left, base]{\color{textcolor}\sffamily\fontsize{6.000000}{7.200000}\selectfont 20\%}%
\end{pgfscope}%
\begin{pgfscope}%
\pgfpathrectangle{\pgfqpoint{0.536528in}{0.633889in}}{\pgfqpoint{2.835476in}{1.604444in}}%
\pgfusepath{clip}%
\pgfsetbuttcap%
\pgfsetroundjoin%
\pgfsetlinewidth{0.401500pt}%
\definecolor{currentstroke}{rgb}{0.647059,0.647059,0.647059}%
\pgfsetstrokecolor{currentstroke}%
\pgfsetdash{}{0pt}%
\pgfpathmoveto{\pgfqpoint{0.536528in}{1.234301in}}%
\pgfpathlineto{\pgfqpoint{3.372004in}{1.234301in}}%
\pgfusepath{stroke}%
\end{pgfscope}%
\begin{pgfscope}%
\pgfsetbuttcap%
\pgfsetroundjoin%
\definecolor{currentfill}{rgb}{0.000000,0.000000,0.000000}%
\pgfsetfillcolor{currentfill}%
\pgfsetlinewidth{0.401500pt}%
\definecolor{currentstroke}{rgb}{0.000000,0.000000,0.000000}%
\pgfsetstrokecolor{currentstroke}%
\pgfsetdash{}{0pt}%
\pgfsys@defobject{currentmarker}{\pgfqpoint{-0.048611in}{0.000000in}}{\pgfqpoint{-0.000000in}{0.000000in}}{%
\pgfpathmoveto{\pgfqpoint{-0.000000in}{0.000000in}}%
\pgfpathlineto{\pgfqpoint{-0.048611in}{0.000000in}}%
\pgfusepath{stroke,fill}%
}%
\begin{pgfscope}%
\pgfsys@transformshift{0.536528in}{1.234301in}%
\pgfsys@useobject{currentmarker}{}%
\end{pgfscope}%
\end{pgfscope}%
\begin{pgfscope}%
\definecolor{textcolor}{rgb}{0.000000,0.000000,0.000000}%
\pgfsetstrokecolor{textcolor}%
\pgfsetfillcolor{textcolor}%
\pgftext[x=0.276977in, y=1.205365in, left, base]{\color{textcolor}\sffamily\fontsize{6.000000}{7.200000}\selectfont 30\%}%
\end{pgfscope}%
\begin{pgfscope}%
\pgfpathrectangle{\pgfqpoint{0.536528in}{0.633889in}}{\pgfqpoint{2.835476in}{1.604444in}}%
\pgfusepath{clip}%
\pgfsetbuttcap%
\pgfsetroundjoin%
\pgfsetlinewidth{0.401500pt}%
\definecolor{currentstroke}{rgb}{0.647059,0.647059,0.647059}%
\pgfsetstrokecolor{currentstroke}%
\pgfsetdash{}{0pt}%
\pgfpathmoveto{\pgfqpoint{0.536528in}{1.435107in}}%
\pgfpathlineto{\pgfqpoint{3.372004in}{1.435107in}}%
\pgfusepath{stroke}%
\end{pgfscope}%
\begin{pgfscope}%
\pgfsetbuttcap%
\pgfsetroundjoin%
\definecolor{currentfill}{rgb}{0.000000,0.000000,0.000000}%
\pgfsetfillcolor{currentfill}%
\pgfsetlinewidth{0.401500pt}%
\definecolor{currentstroke}{rgb}{0.000000,0.000000,0.000000}%
\pgfsetstrokecolor{currentstroke}%
\pgfsetdash{}{0pt}%
\pgfsys@defobject{currentmarker}{\pgfqpoint{-0.048611in}{0.000000in}}{\pgfqpoint{-0.000000in}{0.000000in}}{%
\pgfpathmoveto{\pgfqpoint{-0.000000in}{0.000000in}}%
\pgfpathlineto{\pgfqpoint{-0.048611in}{0.000000in}}%
\pgfusepath{stroke,fill}%
}%
\begin{pgfscope}%
\pgfsys@transformshift{0.536528in}{1.435107in}%
\pgfsys@useobject{currentmarker}{}%
\end{pgfscope}%
\end{pgfscope}%
\begin{pgfscope}%
\definecolor{textcolor}{rgb}{0.000000,0.000000,0.000000}%
\pgfsetstrokecolor{textcolor}%
\pgfsetfillcolor{textcolor}%
\pgftext[x=0.276977in, y=1.406172in, left, base]{\color{textcolor}\sffamily\fontsize{6.000000}{7.200000}\selectfont 40\%}%
\end{pgfscope}%
\begin{pgfscope}%
\pgfpathrectangle{\pgfqpoint{0.536528in}{0.633889in}}{\pgfqpoint{2.835476in}{1.604444in}}%
\pgfusepath{clip}%
\pgfsetbuttcap%
\pgfsetroundjoin%
\pgfsetlinewidth{0.401500pt}%
\definecolor{currentstroke}{rgb}{0.647059,0.647059,0.647059}%
\pgfsetstrokecolor{currentstroke}%
\pgfsetdash{}{0pt}%
\pgfpathmoveto{\pgfqpoint{0.536528in}{1.635914in}}%
\pgfpathlineto{\pgfqpoint{3.372004in}{1.635914in}}%
\pgfusepath{stroke}%
\end{pgfscope}%
\begin{pgfscope}%
\pgfsetbuttcap%
\pgfsetroundjoin%
\definecolor{currentfill}{rgb}{0.000000,0.000000,0.000000}%
\pgfsetfillcolor{currentfill}%
\pgfsetlinewidth{0.401500pt}%
\definecolor{currentstroke}{rgb}{0.000000,0.000000,0.000000}%
\pgfsetstrokecolor{currentstroke}%
\pgfsetdash{}{0pt}%
\pgfsys@defobject{currentmarker}{\pgfqpoint{-0.048611in}{0.000000in}}{\pgfqpoint{-0.000000in}{0.000000in}}{%
\pgfpathmoveto{\pgfqpoint{-0.000000in}{0.000000in}}%
\pgfpathlineto{\pgfqpoint{-0.048611in}{0.000000in}}%
\pgfusepath{stroke,fill}%
}%
\begin{pgfscope}%
\pgfsys@transformshift{0.536528in}{1.635914in}%
\pgfsys@useobject{currentmarker}{}%
\end{pgfscope}%
\end{pgfscope}%
\begin{pgfscope}%
\definecolor{textcolor}{rgb}{0.000000,0.000000,0.000000}%
\pgfsetstrokecolor{textcolor}%
\pgfsetfillcolor{textcolor}%
\pgftext[x=0.276977in, y=1.606979in, left, base]{\color{textcolor}\sffamily\fontsize{6.000000}{7.200000}\selectfont 50\%}%
\end{pgfscope}%
\begin{pgfscope}%
\pgfpathrectangle{\pgfqpoint{0.536528in}{0.633889in}}{\pgfqpoint{2.835476in}{1.604444in}}%
\pgfusepath{clip}%
\pgfsetbuttcap%
\pgfsetroundjoin%
\pgfsetlinewidth{0.401500pt}%
\definecolor{currentstroke}{rgb}{0.647059,0.647059,0.647059}%
\pgfsetstrokecolor{currentstroke}%
\pgfsetdash{}{0pt}%
\pgfpathmoveto{\pgfqpoint{0.536528in}{1.836720in}}%
\pgfpathlineto{\pgfqpoint{3.372004in}{1.836720in}}%
\pgfusepath{stroke}%
\end{pgfscope}%
\begin{pgfscope}%
\pgfsetbuttcap%
\pgfsetroundjoin%
\definecolor{currentfill}{rgb}{0.000000,0.000000,0.000000}%
\pgfsetfillcolor{currentfill}%
\pgfsetlinewidth{0.401500pt}%
\definecolor{currentstroke}{rgb}{0.000000,0.000000,0.000000}%
\pgfsetstrokecolor{currentstroke}%
\pgfsetdash{}{0pt}%
\pgfsys@defobject{currentmarker}{\pgfqpoint{-0.048611in}{0.000000in}}{\pgfqpoint{-0.000000in}{0.000000in}}{%
\pgfpathmoveto{\pgfqpoint{-0.000000in}{0.000000in}}%
\pgfpathlineto{\pgfqpoint{-0.048611in}{0.000000in}}%
\pgfusepath{stroke,fill}%
}%
\begin{pgfscope}%
\pgfsys@transformshift{0.536528in}{1.836720in}%
\pgfsys@useobject{currentmarker}{}%
\end{pgfscope}%
\end{pgfscope}%
\begin{pgfscope}%
\definecolor{textcolor}{rgb}{0.000000,0.000000,0.000000}%
\pgfsetstrokecolor{textcolor}%
\pgfsetfillcolor{textcolor}%
\pgftext[x=0.276977in, y=1.807785in, left, base]{\color{textcolor}\sffamily\fontsize{6.000000}{7.200000}\selectfont 60\%}%
\end{pgfscope}%
\begin{pgfscope}%
\pgfpathrectangle{\pgfqpoint{0.536528in}{0.633889in}}{\pgfqpoint{2.835476in}{1.604444in}}%
\pgfusepath{clip}%
\pgfsetbuttcap%
\pgfsetroundjoin%
\pgfsetlinewidth{0.401500pt}%
\definecolor{currentstroke}{rgb}{0.647059,0.647059,0.647059}%
\pgfsetstrokecolor{currentstroke}%
\pgfsetdash{}{0pt}%
\pgfpathmoveto{\pgfqpoint{0.536528in}{2.037527in}}%
\pgfpathlineto{\pgfqpoint{3.372004in}{2.037527in}}%
\pgfusepath{stroke}%
\end{pgfscope}%
\begin{pgfscope}%
\pgfsetbuttcap%
\pgfsetroundjoin%
\definecolor{currentfill}{rgb}{0.000000,0.000000,0.000000}%
\pgfsetfillcolor{currentfill}%
\pgfsetlinewidth{0.401500pt}%
\definecolor{currentstroke}{rgb}{0.000000,0.000000,0.000000}%
\pgfsetstrokecolor{currentstroke}%
\pgfsetdash{}{0pt}%
\pgfsys@defobject{currentmarker}{\pgfqpoint{-0.048611in}{0.000000in}}{\pgfqpoint{-0.000000in}{0.000000in}}{%
\pgfpathmoveto{\pgfqpoint{-0.000000in}{0.000000in}}%
\pgfpathlineto{\pgfqpoint{-0.048611in}{0.000000in}}%
\pgfusepath{stroke,fill}%
}%
\begin{pgfscope}%
\pgfsys@transformshift{0.536528in}{2.037527in}%
\pgfsys@useobject{currentmarker}{}%
\end{pgfscope}%
\end{pgfscope}%
\begin{pgfscope}%
\definecolor{textcolor}{rgb}{0.000000,0.000000,0.000000}%
\pgfsetstrokecolor{textcolor}%
\pgfsetfillcolor{textcolor}%
\pgftext[x=0.276977in, y=2.008592in, left, base]{\color{textcolor}\sffamily\fontsize{6.000000}{7.200000}\selectfont 70\%}%
\end{pgfscope}%
\begin{pgfscope}%
\pgfpathrectangle{\pgfqpoint{0.536528in}{0.633889in}}{\pgfqpoint{2.835476in}{1.604444in}}%
\pgfusepath{clip}%
\pgfsetbuttcap%
\pgfsetroundjoin%
\pgfsetlinewidth{0.401500pt}%
\definecolor{currentstroke}{rgb}{0.647059,0.647059,0.647059}%
\pgfsetstrokecolor{currentstroke}%
\pgfsetdash{}{0pt}%
\pgfpathmoveto{\pgfqpoint{0.536528in}{2.238333in}}%
\pgfpathlineto{\pgfqpoint{3.372004in}{2.238333in}}%
\pgfusepath{stroke}%
\end{pgfscope}%
\begin{pgfscope}%
\pgfsetbuttcap%
\pgfsetroundjoin%
\definecolor{currentfill}{rgb}{0.000000,0.000000,0.000000}%
\pgfsetfillcolor{currentfill}%
\pgfsetlinewidth{0.401500pt}%
\definecolor{currentstroke}{rgb}{0.000000,0.000000,0.000000}%
\pgfsetstrokecolor{currentstroke}%
\pgfsetdash{}{0pt}%
\pgfsys@defobject{currentmarker}{\pgfqpoint{-0.048611in}{0.000000in}}{\pgfqpoint{-0.000000in}{0.000000in}}{%
\pgfpathmoveto{\pgfqpoint{-0.000000in}{0.000000in}}%
\pgfpathlineto{\pgfqpoint{-0.048611in}{0.000000in}}%
\pgfusepath{stroke,fill}%
}%
\begin{pgfscope}%
\pgfsys@transformshift{0.536528in}{2.238333in}%
\pgfsys@useobject{currentmarker}{}%
\end{pgfscope}%
\end{pgfscope}%
\begin{pgfscope}%
\definecolor{textcolor}{rgb}{0.000000,0.000000,0.000000}%
\pgfsetstrokecolor{textcolor}%
\pgfsetfillcolor{textcolor}%
\pgftext[x=0.276977in, y=2.209398in, left, base]{\color{textcolor}\sffamily\fontsize{6.000000}{7.200000}\selectfont 80\%}%
\end{pgfscope}%
\begin{pgfscope}%
\definecolor{textcolor}{rgb}{0.000000,0.000000,0.000000}%
\pgfsetstrokecolor{textcolor}%
\pgfsetfillcolor{textcolor}%
\pgftext[x=0.221421in,y=1.436111in,,bottom,rotate=90.000000]{\color{textcolor}\sffamily\fontsize{7.000000}{8.400000}\selectfont Sample Accuracy (Session 2)}%
\end{pgfscope}%
\begin{pgfscope}%
\pgfpathrectangle{\pgfqpoint{0.536528in}{0.633889in}}{\pgfqpoint{2.835476in}{1.604444in}}%
\pgfusepath{clip}%
\pgfsetbuttcap%
\pgfsetmiterjoin%
\definecolor{currentfill}{rgb}{1.000000,0.650980,0.000000}%
\pgfsetfillcolor{currentfill}%
\pgfsetlinewidth{0.000000pt}%
\definecolor{currentstroke}{rgb}{0.000000,0.000000,0.000000}%
\pgfsetstrokecolor{currentstroke}%
\pgfsetstrokeopacity{0.000000}%
\pgfsetdash{}{0pt}%
\pgfpathmoveto{\pgfqpoint{0.772817in}{0.631881in}}%
\pgfpathlineto{\pgfqpoint{0.941596in}{0.631881in}}%
\pgfpathlineto{\pgfqpoint{0.941596in}{1.796559in}}%
\pgfpathlineto{\pgfqpoint{0.772817in}{1.796559in}}%
\pgfpathlineto{\pgfqpoint{0.772817in}{0.631881in}}%
\pgfpathclose%
\pgfusepath{fill}%
\end{pgfscope}%
\begin{pgfscope}%
\pgfpathrectangle{\pgfqpoint{0.536528in}{0.633889in}}{\pgfqpoint{2.835476in}{1.604444in}}%
\pgfusepath{clip}%
\pgfsetbuttcap%
\pgfsetmiterjoin%
\definecolor{currentfill}{rgb}{1.000000,0.650980,0.000000}%
\pgfsetfillcolor{currentfill}%
\pgfsetlinewidth{0.000000pt}%
\definecolor{currentstroke}{rgb}{0.000000,0.000000,0.000000}%
\pgfsetstrokecolor{currentstroke}%
\pgfsetstrokeopacity{0.000000}%
\pgfsetdash{}{0pt}%
\pgfpathmoveto{\pgfqpoint{1.447931in}{0.631881in}}%
\pgfpathlineto{\pgfqpoint{1.616709in}{0.631881in}}%
\pgfpathlineto{\pgfqpoint{1.616709in}{1.937123in}}%
\pgfpathlineto{\pgfqpoint{1.447931in}{1.937123in}}%
\pgfpathlineto{\pgfqpoint{1.447931in}{0.631881in}}%
\pgfpathclose%
\pgfusepath{fill}%
\end{pgfscope}%
\begin{pgfscope}%
\pgfpathrectangle{\pgfqpoint{0.536528in}{0.633889in}}{\pgfqpoint{2.835476in}{1.604444in}}%
\pgfusepath{clip}%
\pgfsetbuttcap%
\pgfsetmiterjoin%
\definecolor{currentfill}{rgb}{1.000000,0.650980,0.000000}%
\pgfsetfillcolor{currentfill}%
\pgfsetlinewidth{0.000000pt}%
\definecolor{currentstroke}{rgb}{0.000000,0.000000,0.000000}%
\pgfsetstrokecolor{currentstroke}%
\pgfsetstrokeopacity{0.000000}%
\pgfsetdash{}{0pt}%
\pgfpathmoveto{\pgfqpoint{2.123044in}{0.631881in}}%
\pgfpathlineto{\pgfqpoint{2.291823in}{0.631881in}}%
\pgfpathlineto{\pgfqpoint{2.291823in}{1.997365in}}%
\pgfpathlineto{\pgfqpoint{2.123044in}{1.997365in}}%
\pgfpathlineto{\pgfqpoint{2.123044in}{0.631881in}}%
\pgfpathclose%
\pgfusepath{fill}%
\end{pgfscope}%
\begin{pgfscope}%
\pgfpathrectangle{\pgfqpoint{0.536528in}{0.633889in}}{\pgfqpoint{2.835476in}{1.604444in}}%
\pgfusepath{clip}%
\pgfsetbuttcap%
\pgfsetmiterjoin%
\definecolor{currentfill}{rgb}{1.000000,0.650980,0.000000}%
\pgfsetfillcolor{currentfill}%
\pgfsetlinewidth{0.000000pt}%
\definecolor{currentstroke}{rgb}{0.000000,0.000000,0.000000}%
\pgfsetstrokecolor{currentstroke}%
\pgfsetstrokeopacity{0.000000}%
\pgfsetdash{}{0pt}%
\pgfpathmoveto{\pgfqpoint{2.798158in}{0.631881in}}%
\pgfpathlineto{\pgfqpoint{2.966936in}{0.631881in}}%
\pgfpathlineto{\pgfqpoint{2.966936in}{1.997365in}}%
\pgfpathlineto{\pgfqpoint{2.798158in}{1.997365in}}%
\pgfpathlineto{\pgfqpoint{2.798158in}{0.631881in}}%
\pgfpathclose%
\pgfusepath{fill}%
\end{pgfscope}%
\begin{pgfscope}%
\pgfpathrectangle{\pgfqpoint{0.536528in}{0.633889in}}{\pgfqpoint{2.835476in}{1.604444in}}%
\pgfusepath{clip}%
\pgfsetbuttcap%
\pgfsetmiterjoin%
\definecolor{currentfill}{rgb}{0.666667,0.666667,0.666667}%
\pgfsetfillcolor{currentfill}%
\pgfsetlinewidth{0.000000pt}%
\definecolor{currentstroke}{rgb}{0.000000,0.000000,0.000000}%
\pgfsetstrokecolor{currentstroke}%
\pgfsetstrokeopacity{0.000000}%
\pgfsetdash{}{0pt}%
\pgfpathmoveto{\pgfqpoint{0.941596in}{0.631881in}}%
\pgfpathlineto{\pgfqpoint{1.110374in}{0.631881in}}%
\pgfpathlineto{\pgfqpoint{1.110374in}{1.535510in}}%
\pgfpathlineto{\pgfqpoint{0.941596in}{1.535510in}}%
\pgfpathlineto{\pgfqpoint{0.941596in}{0.631881in}}%
\pgfpathclose%
\pgfusepath{fill}%
\end{pgfscope}%
\begin{pgfscope}%
\pgfpathrectangle{\pgfqpoint{0.536528in}{0.633889in}}{\pgfqpoint{2.835476in}{1.604444in}}%
\pgfusepath{clip}%
\pgfsetbuttcap%
\pgfsetmiterjoin%
\definecolor{currentfill}{rgb}{0.666667,0.666667,0.666667}%
\pgfsetfillcolor{currentfill}%
\pgfsetlinewidth{0.000000pt}%
\definecolor{currentstroke}{rgb}{0.000000,0.000000,0.000000}%
\pgfsetstrokecolor{currentstroke}%
\pgfsetstrokeopacity{0.000000}%
\pgfsetdash{}{0pt}%
\pgfpathmoveto{\pgfqpoint{1.616709in}{0.631881in}}%
\pgfpathlineto{\pgfqpoint{1.785488in}{0.631881in}}%
\pgfpathlineto{\pgfqpoint{1.785488in}{1.796559in}}%
\pgfpathlineto{\pgfqpoint{1.616709in}{1.796559in}}%
\pgfpathlineto{\pgfqpoint{1.616709in}{0.631881in}}%
\pgfpathclose%
\pgfusepath{fill}%
\end{pgfscope}%
\begin{pgfscope}%
\pgfpathrectangle{\pgfqpoint{0.536528in}{0.633889in}}{\pgfqpoint{2.835476in}{1.604444in}}%
\pgfusepath{clip}%
\pgfsetbuttcap%
\pgfsetmiterjoin%
\definecolor{currentfill}{rgb}{0.666667,0.666667,0.666667}%
\pgfsetfillcolor{currentfill}%
\pgfsetlinewidth{0.000000pt}%
\definecolor{currentstroke}{rgb}{0.000000,0.000000,0.000000}%
\pgfsetstrokecolor{currentstroke}%
\pgfsetstrokeopacity{0.000000}%
\pgfsetdash{}{0pt}%
\pgfpathmoveto{\pgfqpoint{2.291823in}{0.631881in}}%
\pgfpathlineto{\pgfqpoint{2.460601in}{0.631881in}}%
\pgfpathlineto{\pgfqpoint{2.460601in}{1.937123in}}%
\pgfpathlineto{\pgfqpoint{2.291823in}{1.937123in}}%
\pgfpathlineto{\pgfqpoint{2.291823in}{0.631881in}}%
\pgfpathclose%
\pgfusepath{fill}%
\end{pgfscope}%
\begin{pgfscope}%
\pgfpathrectangle{\pgfqpoint{0.536528in}{0.633889in}}{\pgfqpoint{2.835476in}{1.604444in}}%
\pgfusepath{clip}%
\pgfsetbuttcap%
\pgfsetmiterjoin%
\definecolor{currentfill}{rgb}{0.666667,0.666667,0.666667}%
\pgfsetfillcolor{currentfill}%
\pgfsetlinewidth{0.000000pt}%
\definecolor{currentstroke}{rgb}{0.000000,0.000000,0.000000}%
\pgfsetstrokecolor{currentstroke}%
\pgfsetstrokeopacity{0.000000}%
\pgfsetdash{}{0pt}%
\pgfpathmoveto{\pgfqpoint{2.966936in}{0.631881in}}%
\pgfpathlineto{\pgfqpoint{3.135714in}{0.631881in}}%
\pgfpathlineto{\pgfqpoint{3.135714in}{1.756398in}}%
\pgfpathlineto{\pgfqpoint{2.966936in}{1.756398in}}%
\pgfpathlineto{\pgfqpoint{2.966936in}{0.631881in}}%
\pgfpathclose%
\pgfusepath{fill}%
\end{pgfscope}%
\begin{pgfscope}%
\pgfpathrectangle{\pgfqpoint{0.536528in}{0.633889in}}{\pgfqpoint{2.835476in}{1.604444in}}%
\pgfusepath{clip}%
\pgfsetbuttcap%
\pgfsetroundjoin%
\pgfsetlinewidth{1.003750pt}%
\definecolor{currentstroke}{rgb}{0.000000,0.000000,0.000000}%
\pgfsetstrokecolor{currentstroke}%
\pgfsetdash{}{0pt}%
\pgfpathmoveto{\pgfqpoint{0.860582in}{1.709379in}}%
\pgfpathlineto{\pgfqpoint{0.860582in}{1.900391in}}%
\pgfusepath{stroke}%
\end{pgfscope}%
\begin{pgfscope}%
\pgfpathrectangle{\pgfqpoint{0.536528in}{0.633889in}}{\pgfqpoint{2.835476in}{1.604444in}}%
\pgfusepath{clip}%
\pgfsetbuttcap%
\pgfsetroundjoin%
\definecolor{currentfill}{rgb}{0.000000,0.000000,0.000000}%
\pgfsetfillcolor{currentfill}%
\pgfsetlinewidth{1.003750pt}%
\definecolor{currentstroke}{rgb}{0.000000,0.000000,0.000000}%
\pgfsetstrokecolor{currentstroke}%
\pgfsetdash{}{0pt}%
\pgfsys@defobject{currentmarker}{\pgfqpoint{-0.027778in}{-0.000000in}}{\pgfqpoint{0.027778in}{0.000000in}}{%
\pgfpathmoveto{\pgfqpoint{0.027778in}{-0.000000in}}%
\pgfpathlineto{\pgfqpoint{-0.027778in}{0.000000in}}%
\pgfusepath{stroke,fill}%
}%
\begin{pgfscope}%
\pgfsys@transformshift{0.860582in}{1.709379in}%
\pgfsys@useobject{currentmarker}{}%
\end{pgfscope}%
\end{pgfscope}%
\begin{pgfscope}%
\pgfpathrectangle{\pgfqpoint{0.536528in}{0.633889in}}{\pgfqpoint{2.835476in}{1.604444in}}%
\pgfusepath{clip}%
\pgfsetbuttcap%
\pgfsetroundjoin%
\definecolor{currentfill}{rgb}{0.000000,0.000000,0.000000}%
\pgfsetfillcolor{currentfill}%
\pgfsetlinewidth{1.003750pt}%
\definecolor{currentstroke}{rgb}{0.000000,0.000000,0.000000}%
\pgfsetstrokecolor{currentstroke}%
\pgfsetdash{}{0pt}%
\pgfsys@defobject{currentmarker}{\pgfqpoint{-0.027778in}{-0.000000in}}{\pgfqpoint{0.027778in}{0.000000in}}{%
\pgfpathmoveto{\pgfqpoint{0.027778in}{-0.000000in}}%
\pgfpathlineto{\pgfqpoint{-0.027778in}{0.000000in}}%
\pgfusepath{stroke,fill}%
}%
\begin{pgfscope}%
\pgfsys@transformshift{0.860582in}{1.900391in}%
\pgfsys@useobject{currentmarker}{}%
\end{pgfscope}%
\end{pgfscope}%
\begin{pgfscope}%
\pgfpathrectangle{\pgfqpoint{0.536528in}{0.633889in}}{\pgfqpoint{2.835476in}{1.604444in}}%
\pgfusepath{clip}%
\pgfsetbuttcap%
\pgfsetroundjoin%
\pgfsetlinewidth{1.003750pt}%
\definecolor{currentstroke}{rgb}{0.000000,0.000000,0.000000}%
\pgfsetstrokecolor{currentstroke}%
\pgfsetdash{}{0pt}%
\pgfpathmoveto{\pgfqpoint{1.029361in}{1.430209in}}%
\pgfpathlineto{\pgfqpoint{1.029361in}{1.626118in}}%
\pgfusepath{stroke}%
\end{pgfscope}%
\begin{pgfscope}%
\pgfpathrectangle{\pgfqpoint{0.536528in}{0.633889in}}{\pgfqpoint{2.835476in}{1.604444in}}%
\pgfusepath{clip}%
\pgfsetbuttcap%
\pgfsetroundjoin%
\definecolor{currentfill}{rgb}{0.000000,0.000000,0.000000}%
\pgfsetfillcolor{currentfill}%
\pgfsetlinewidth{1.003750pt}%
\definecolor{currentstroke}{rgb}{0.000000,0.000000,0.000000}%
\pgfsetstrokecolor{currentstroke}%
\pgfsetdash{}{0pt}%
\pgfsys@defobject{currentmarker}{\pgfqpoint{-0.027778in}{-0.000000in}}{\pgfqpoint{0.027778in}{0.000000in}}{%
\pgfpathmoveto{\pgfqpoint{0.027778in}{-0.000000in}}%
\pgfpathlineto{\pgfqpoint{-0.027778in}{0.000000in}}%
\pgfusepath{stroke,fill}%
}%
\begin{pgfscope}%
\pgfsys@transformshift{1.029361in}{1.430209in}%
\pgfsys@useobject{currentmarker}{}%
\end{pgfscope}%
\end{pgfscope}%
\begin{pgfscope}%
\pgfpathrectangle{\pgfqpoint{0.536528in}{0.633889in}}{\pgfqpoint{2.835476in}{1.604444in}}%
\pgfusepath{clip}%
\pgfsetbuttcap%
\pgfsetroundjoin%
\definecolor{currentfill}{rgb}{0.000000,0.000000,0.000000}%
\pgfsetfillcolor{currentfill}%
\pgfsetlinewidth{1.003750pt}%
\definecolor{currentstroke}{rgb}{0.000000,0.000000,0.000000}%
\pgfsetstrokecolor{currentstroke}%
\pgfsetdash{}{0pt}%
\pgfsys@defobject{currentmarker}{\pgfqpoint{-0.027778in}{-0.000000in}}{\pgfqpoint{0.027778in}{0.000000in}}{%
\pgfpathmoveto{\pgfqpoint{0.027778in}{-0.000000in}}%
\pgfpathlineto{\pgfqpoint{-0.027778in}{0.000000in}}%
\pgfusepath{stroke,fill}%
}%
\begin{pgfscope}%
\pgfsys@transformshift{1.029361in}{1.626118in}%
\pgfsys@useobject{currentmarker}{}%
\end{pgfscope}%
\end{pgfscope}%
\begin{pgfscope}%
\pgfpathrectangle{\pgfqpoint{0.536528in}{0.633889in}}{\pgfqpoint{2.835476in}{1.604444in}}%
\pgfusepath{clip}%
\pgfsetbuttcap%
\pgfsetroundjoin%
\pgfsetlinewidth{1.003750pt}%
\definecolor{currentstroke}{rgb}{0.000000,0.000000,0.000000}%
\pgfsetstrokecolor{currentstroke}%
\pgfsetdash{}{0pt}%
\pgfpathmoveto{\pgfqpoint{1.535696in}{1.846516in}}%
\pgfpathlineto{\pgfqpoint{1.535696in}{2.032629in}}%
\pgfusepath{stroke}%
\end{pgfscope}%
\begin{pgfscope}%
\pgfpathrectangle{\pgfqpoint{0.536528in}{0.633889in}}{\pgfqpoint{2.835476in}{1.604444in}}%
\pgfusepath{clip}%
\pgfsetbuttcap%
\pgfsetroundjoin%
\definecolor{currentfill}{rgb}{0.000000,0.000000,0.000000}%
\pgfsetfillcolor{currentfill}%
\pgfsetlinewidth{1.003750pt}%
\definecolor{currentstroke}{rgb}{0.000000,0.000000,0.000000}%
\pgfsetstrokecolor{currentstroke}%
\pgfsetdash{}{0pt}%
\pgfsys@defobject{currentmarker}{\pgfqpoint{-0.027778in}{-0.000000in}}{\pgfqpoint{0.027778in}{0.000000in}}{%
\pgfpathmoveto{\pgfqpoint{0.027778in}{-0.000000in}}%
\pgfpathlineto{\pgfqpoint{-0.027778in}{0.000000in}}%
\pgfusepath{stroke,fill}%
}%
\begin{pgfscope}%
\pgfsys@transformshift{1.535696in}{1.846516in}%
\pgfsys@useobject{currentmarker}{}%
\end{pgfscope}%
\end{pgfscope}%
\begin{pgfscope}%
\pgfpathrectangle{\pgfqpoint{0.536528in}{0.633889in}}{\pgfqpoint{2.835476in}{1.604444in}}%
\pgfusepath{clip}%
\pgfsetbuttcap%
\pgfsetroundjoin%
\definecolor{currentfill}{rgb}{0.000000,0.000000,0.000000}%
\pgfsetfillcolor{currentfill}%
\pgfsetlinewidth{1.003750pt}%
\definecolor{currentstroke}{rgb}{0.000000,0.000000,0.000000}%
\pgfsetstrokecolor{currentstroke}%
\pgfsetdash{}{0pt}%
\pgfsys@defobject{currentmarker}{\pgfqpoint{-0.027778in}{-0.000000in}}{\pgfqpoint{0.027778in}{0.000000in}}{%
\pgfpathmoveto{\pgfqpoint{0.027778in}{-0.000000in}}%
\pgfpathlineto{\pgfqpoint{-0.027778in}{0.000000in}}%
\pgfusepath{stroke,fill}%
}%
\begin{pgfscope}%
\pgfsys@transformshift{1.535696in}{2.032629in}%
\pgfsys@useobject{currentmarker}{}%
\end{pgfscope}%
\end{pgfscope}%
\begin{pgfscope}%
\pgfpathrectangle{\pgfqpoint{0.536528in}{0.633889in}}{\pgfqpoint{2.835476in}{1.604444in}}%
\pgfusepath{clip}%
\pgfsetbuttcap%
\pgfsetroundjoin%
\pgfsetlinewidth{1.003750pt}%
\definecolor{currentstroke}{rgb}{0.000000,0.000000,0.000000}%
\pgfsetstrokecolor{currentstroke}%
\pgfsetdash{}{0pt}%
\pgfpathmoveto{\pgfqpoint{1.704474in}{1.699584in}}%
\pgfpathlineto{\pgfqpoint{1.704474in}{1.890595in}}%
\pgfusepath{stroke}%
\end{pgfscope}%
\begin{pgfscope}%
\pgfpathrectangle{\pgfqpoint{0.536528in}{0.633889in}}{\pgfqpoint{2.835476in}{1.604444in}}%
\pgfusepath{clip}%
\pgfsetbuttcap%
\pgfsetroundjoin%
\definecolor{currentfill}{rgb}{0.000000,0.000000,0.000000}%
\pgfsetfillcolor{currentfill}%
\pgfsetlinewidth{1.003750pt}%
\definecolor{currentstroke}{rgb}{0.000000,0.000000,0.000000}%
\pgfsetstrokecolor{currentstroke}%
\pgfsetdash{}{0pt}%
\pgfsys@defobject{currentmarker}{\pgfqpoint{-0.027778in}{-0.000000in}}{\pgfqpoint{0.027778in}{0.000000in}}{%
\pgfpathmoveto{\pgfqpoint{0.027778in}{-0.000000in}}%
\pgfpathlineto{\pgfqpoint{-0.027778in}{0.000000in}}%
\pgfusepath{stroke,fill}%
}%
\begin{pgfscope}%
\pgfsys@transformshift{1.704474in}{1.699584in}%
\pgfsys@useobject{currentmarker}{}%
\end{pgfscope}%
\end{pgfscope}%
\begin{pgfscope}%
\pgfpathrectangle{\pgfqpoint{0.536528in}{0.633889in}}{\pgfqpoint{2.835476in}{1.604444in}}%
\pgfusepath{clip}%
\pgfsetbuttcap%
\pgfsetroundjoin%
\definecolor{currentfill}{rgb}{0.000000,0.000000,0.000000}%
\pgfsetfillcolor{currentfill}%
\pgfsetlinewidth{1.003750pt}%
\definecolor{currentstroke}{rgb}{0.000000,0.000000,0.000000}%
\pgfsetstrokecolor{currentstroke}%
\pgfsetdash{}{0pt}%
\pgfsys@defobject{currentmarker}{\pgfqpoint{-0.027778in}{-0.000000in}}{\pgfqpoint{0.027778in}{0.000000in}}{%
\pgfpathmoveto{\pgfqpoint{0.027778in}{-0.000000in}}%
\pgfpathlineto{\pgfqpoint{-0.027778in}{0.000000in}}%
\pgfusepath{stroke,fill}%
}%
\begin{pgfscope}%
\pgfsys@transformshift{1.704474in}{1.890595in}%
\pgfsys@useobject{currentmarker}{}%
\end{pgfscope}%
\end{pgfscope}%
\begin{pgfscope}%
\pgfpathrectangle{\pgfqpoint{0.536528in}{0.633889in}}{\pgfqpoint{2.835476in}{1.604444in}}%
\pgfusepath{clip}%
\pgfsetbuttcap%
\pgfsetroundjoin%
\pgfsetlinewidth{1.003750pt}%
\definecolor{currentstroke}{rgb}{0.000000,0.000000,0.000000}%
\pgfsetstrokecolor{currentstroke}%
\pgfsetdash{}{0pt}%
\pgfpathmoveto{\pgfqpoint{2.210809in}{1.905288in}}%
\pgfpathlineto{\pgfqpoint{2.210809in}{2.086504in}}%
\pgfusepath{stroke}%
\end{pgfscope}%
\begin{pgfscope}%
\pgfpathrectangle{\pgfqpoint{0.536528in}{0.633889in}}{\pgfqpoint{2.835476in}{1.604444in}}%
\pgfusepath{clip}%
\pgfsetbuttcap%
\pgfsetroundjoin%
\definecolor{currentfill}{rgb}{0.000000,0.000000,0.000000}%
\pgfsetfillcolor{currentfill}%
\pgfsetlinewidth{1.003750pt}%
\definecolor{currentstroke}{rgb}{0.000000,0.000000,0.000000}%
\pgfsetstrokecolor{currentstroke}%
\pgfsetdash{}{0pt}%
\pgfsys@defobject{currentmarker}{\pgfqpoint{-0.027778in}{-0.000000in}}{\pgfqpoint{0.027778in}{0.000000in}}{%
\pgfpathmoveto{\pgfqpoint{0.027778in}{-0.000000in}}%
\pgfpathlineto{\pgfqpoint{-0.027778in}{0.000000in}}%
\pgfusepath{stroke,fill}%
}%
\begin{pgfscope}%
\pgfsys@transformshift{2.210809in}{1.905288in}%
\pgfsys@useobject{currentmarker}{}%
\end{pgfscope}%
\end{pgfscope}%
\begin{pgfscope}%
\pgfpathrectangle{\pgfqpoint{0.536528in}{0.633889in}}{\pgfqpoint{2.835476in}{1.604444in}}%
\pgfusepath{clip}%
\pgfsetbuttcap%
\pgfsetroundjoin%
\definecolor{currentfill}{rgb}{0.000000,0.000000,0.000000}%
\pgfsetfillcolor{currentfill}%
\pgfsetlinewidth{1.003750pt}%
\definecolor{currentstroke}{rgb}{0.000000,0.000000,0.000000}%
\pgfsetstrokecolor{currentstroke}%
\pgfsetdash{}{0pt}%
\pgfsys@defobject{currentmarker}{\pgfqpoint{-0.027778in}{-0.000000in}}{\pgfqpoint{0.027778in}{0.000000in}}{%
\pgfpathmoveto{\pgfqpoint{0.027778in}{-0.000000in}}%
\pgfpathlineto{\pgfqpoint{-0.027778in}{0.000000in}}%
\pgfusepath{stroke,fill}%
}%
\begin{pgfscope}%
\pgfsys@transformshift{2.210809in}{2.086504in}%
\pgfsys@useobject{currentmarker}{}%
\end{pgfscope}%
\end{pgfscope}%
\begin{pgfscope}%
\pgfpathrectangle{\pgfqpoint{0.536528in}{0.633889in}}{\pgfqpoint{2.835476in}{1.604444in}}%
\pgfusepath{clip}%
\pgfsetbuttcap%
\pgfsetroundjoin%
\pgfsetlinewidth{1.003750pt}%
\definecolor{currentstroke}{rgb}{0.000000,0.000000,0.000000}%
\pgfsetstrokecolor{currentstroke}%
\pgfsetdash{}{0pt}%
\pgfpathmoveto{\pgfqpoint{2.379587in}{1.836720in}}%
\pgfpathlineto{\pgfqpoint{2.379587in}{2.022834in}}%
\pgfusepath{stroke}%
\end{pgfscope}%
\begin{pgfscope}%
\pgfpathrectangle{\pgfqpoint{0.536528in}{0.633889in}}{\pgfqpoint{2.835476in}{1.604444in}}%
\pgfusepath{clip}%
\pgfsetbuttcap%
\pgfsetroundjoin%
\definecolor{currentfill}{rgb}{0.000000,0.000000,0.000000}%
\pgfsetfillcolor{currentfill}%
\pgfsetlinewidth{1.003750pt}%
\definecolor{currentstroke}{rgb}{0.000000,0.000000,0.000000}%
\pgfsetstrokecolor{currentstroke}%
\pgfsetdash{}{0pt}%
\pgfsys@defobject{currentmarker}{\pgfqpoint{-0.027778in}{-0.000000in}}{\pgfqpoint{0.027778in}{0.000000in}}{%
\pgfpathmoveto{\pgfqpoint{0.027778in}{-0.000000in}}%
\pgfpathlineto{\pgfqpoint{-0.027778in}{0.000000in}}%
\pgfusepath{stroke,fill}%
}%
\begin{pgfscope}%
\pgfsys@transformshift{2.379587in}{1.836720in}%
\pgfsys@useobject{currentmarker}{}%
\end{pgfscope}%
\end{pgfscope}%
\begin{pgfscope}%
\pgfpathrectangle{\pgfqpoint{0.536528in}{0.633889in}}{\pgfqpoint{2.835476in}{1.604444in}}%
\pgfusepath{clip}%
\pgfsetbuttcap%
\pgfsetroundjoin%
\definecolor{currentfill}{rgb}{0.000000,0.000000,0.000000}%
\pgfsetfillcolor{currentfill}%
\pgfsetlinewidth{1.003750pt}%
\definecolor{currentstroke}{rgb}{0.000000,0.000000,0.000000}%
\pgfsetstrokecolor{currentstroke}%
\pgfsetdash{}{0pt}%
\pgfsys@defobject{currentmarker}{\pgfqpoint{-0.027778in}{-0.000000in}}{\pgfqpoint{0.027778in}{0.000000in}}{%
\pgfpathmoveto{\pgfqpoint{0.027778in}{-0.000000in}}%
\pgfpathlineto{\pgfqpoint{-0.027778in}{0.000000in}}%
\pgfusepath{stroke,fill}%
}%
\begin{pgfscope}%
\pgfsys@transformshift{2.379587in}{2.022834in}%
\pgfsys@useobject{currentmarker}{}%
\end{pgfscope}%
\end{pgfscope}%
\begin{pgfscope}%
\pgfpathrectangle{\pgfqpoint{0.536528in}{0.633889in}}{\pgfqpoint{2.835476in}{1.604444in}}%
\pgfusepath{clip}%
\pgfsetbuttcap%
\pgfsetroundjoin%
\pgfsetlinewidth{1.003750pt}%
\definecolor{currentstroke}{rgb}{0.000000,0.000000,0.000000}%
\pgfsetstrokecolor{currentstroke}%
\pgfsetdash{}{0pt}%
\pgfpathmoveto{\pgfqpoint{2.885922in}{1.905288in}}%
\pgfpathlineto{\pgfqpoint{2.885922in}{2.086504in}}%
\pgfusepath{stroke}%
\end{pgfscope}%
\begin{pgfscope}%
\pgfpathrectangle{\pgfqpoint{0.536528in}{0.633889in}}{\pgfqpoint{2.835476in}{1.604444in}}%
\pgfusepath{clip}%
\pgfsetbuttcap%
\pgfsetroundjoin%
\definecolor{currentfill}{rgb}{0.000000,0.000000,0.000000}%
\pgfsetfillcolor{currentfill}%
\pgfsetlinewidth{1.003750pt}%
\definecolor{currentstroke}{rgb}{0.000000,0.000000,0.000000}%
\pgfsetstrokecolor{currentstroke}%
\pgfsetdash{}{0pt}%
\pgfsys@defobject{currentmarker}{\pgfqpoint{-0.027778in}{-0.000000in}}{\pgfqpoint{0.027778in}{0.000000in}}{%
\pgfpathmoveto{\pgfqpoint{0.027778in}{-0.000000in}}%
\pgfpathlineto{\pgfqpoint{-0.027778in}{0.000000in}}%
\pgfusepath{stroke,fill}%
}%
\begin{pgfscope}%
\pgfsys@transformshift{2.885922in}{1.905288in}%
\pgfsys@useobject{currentmarker}{}%
\end{pgfscope}%
\end{pgfscope}%
\begin{pgfscope}%
\pgfpathrectangle{\pgfqpoint{0.536528in}{0.633889in}}{\pgfqpoint{2.835476in}{1.604444in}}%
\pgfusepath{clip}%
\pgfsetbuttcap%
\pgfsetroundjoin%
\definecolor{currentfill}{rgb}{0.000000,0.000000,0.000000}%
\pgfsetfillcolor{currentfill}%
\pgfsetlinewidth{1.003750pt}%
\definecolor{currentstroke}{rgb}{0.000000,0.000000,0.000000}%
\pgfsetstrokecolor{currentstroke}%
\pgfsetdash{}{0pt}%
\pgfsys@defobject{currentmarker}{\pgfqpoint{-0.027778in}{-0.000000in}}{\pgfqpoint{0.027778in}{0.000000in}}{%
\pgfpathmoveto{\pgfqpoint{0.027778in}{-0.000000in}}%
\pgfpathlineto{\pgfqpoint{-0.027778in}{0.000000in}}%
\pgfusepath{stroke,fill}%
}%
\begin{pgfscope}%
\pgfsys@transformshift{2.885922in}{2.086504in}%
\pgfsys@useobject{currentmarker}{}%
\end{pgfscope}%
\end{pgfscope}%
\begin{pgfscope}%
\pgfpathrectangle{\pgfqpoint{0.536528in}{0.633889in}}{\pgfqpoint{2.835476in}{1.604444in}}%
\pgfusepath{clip}%
\pgfsetbuttcap%
\pgfsetroundjoin%
\pgfsetlinewidth{1.003750pt}%
\definecolor{currentstroke}{rgb}{0.000000,0.000000,0.000000}%
\pgfsetstrokecolor{currentstroke}%
\pgfsetdash{}{0pt}%
\pgfpathmoveto{\pgfqpoint{3.054701in}{1.694686in}}%
\pgfpathlineto{\pgfqpoint{3.054701in}{1.807334in}}%
\pgfusepath{stroke}%
\end{pgfscope}%
\begin{pgfscope}%
\pgfpathrectangle{\pgfqpoint{0.536528in}{0.633889in}}{\pgfqpoint{2.835476in}{1.604444in}}%
\pgfusepath{clip}%
\pgfsetbuttcap%
\pgfsetroundjoin%
\definecolor{currentfill}{rgb}{0.000000,0.000000,0.000000}%
\pgfsetfillcolor{currentfill}%
\pgfsetlinewidth{1.003750pt}%
\definecolor{currentstroke}{rgb}{0.000000,0.000000,0.000000}%
\pgfsetstrokecolor{currentstroke}%
\pgfsetdash{}{0pt}%
\pgfsys@defobject{currentmarker}{\pgfqpoint{-0.027778in}{-0.000000in}}{\pgfqpoint{0.027778in}{0.000000in}}{%
\pgfpathmoveto{\pgfqpoint{0.027778in}{-0.000000in}}%
\pgfpathlineto{\pgfqpoint{-0.027778in}{0.000000in}}%
\pgfusepath{stroke,fill}%
}%
\begin{pgfscope}%
\pgfsys@transformshift{3.054701in}{1.694686in}%
\pgfsys@useobject{currentmarker}{}%
\end{pgfscope}%
\end{pgfscope}%
\begin{pgfscope}%
\pgfpathrectangle{\pgfqpoint{0.536528in}{0.633889in}}{\pgfqpoint{2.835476in}{1.604444in}}%
\pgfusepath{clip}%
\pgfsetbuttcap%
\pgfsetroundjoin%
\definecolor{currentfill}{rgb}{0.000000,0.000000,0.000000}%
\pgfsetfillcolor{currentfill}%
\pgfsetlinewidth{1.003750pt}%
\definecolor{currentstroke}{rgb}{0.000000,0.000000,0.000000}%
\pgfsetstrokecolor{currentstroke}%
\pgfsetdash{}{0pt}%
\pgfsys@defobject{currentmarker}{\pgfqpoint{-0.027778in}{-0.000000in}}{\pgfqpoint{0.027778in}{0.000000in}}{%
\pgfpathmoveto{\pgfqpoint{0.027778in}{-0.000000in}}%
\pgfpathlineto{\pgfqpoint{-0.027778in}{0.000000in}}%
\pgfusepath{stroke,fill}%
}%
\begin{pgfscope}%
\pgfsys@transformshift{3.054701in}{1.807334in}%
\pgfsys@useobject{currentmarker}{}%
\end{pgfscope}%
\end{pgfscope}%
\begin{pgfscope}%
\pgfpathrectangle{\pgfqpoint{0.536528in}{0.633889in}}{\pgfqpoint{2.835476in}{1.604444in}}%
\pgfusepath{clip}%
\pgfsetbuttcap%
\pgfsetroundjoin%
\pgfsetlinewidth{1.003750pt}%
\definecolor{currentstroke}{rgb}{0.000000,0.000000,0.000000}%
\pgfsetstrokecolor{currentstroke}%
\pgfsetdash{}{0pt}%
\pgfpathmoveto{\pgfqpoint{0.860582in}{1.796559in}}%
\pgfusepath{stroke}%
\end{pgfscope}%
\begin{pgfscope}%
\pgfpathrectangle{\pgfqpoint{0.536528in}{0.633889in}}{\pgfqpoint{2.835476in}{1.604444in}}%
\pgfusepath{clip}%
\pgfsetbuttcap%
\pgfsetroundjoin%
\pgfsetlinewidth{1.003750pt}%
\definecolor{currentstroke}{rgb}{0.000000,0.000000,0.000000}%
\pgfsetstrokecolor{currentstroke}%
\pgfsetdash{}{0pt}%
\pgfpathmoveto{\pgfqpoint{1.029361in}{1.535510in}}%
\pgfusepath{stroke}%
\end{pgfscope}%
\begin{pgfscope}%
\pgfpathrectangle{\pgfqpoint{0.536528in}{0.633889in}}{\pgfqpoint{2.835476in}{1.604444in}}%
\pgfusepath{clip}%
\pgfsetbuttcap%
\pgfsetroundjoin%
\pgfsetlinewidth{1.003750pt}%
\definecolor{currentstroke}{rgb}{0.000000,0.000000,0.000000}%
\pgfsetstrokecolor{currentstroke}%
\pgfsetdash{}{0pt}%
\pgfpathmoveto{\pgfqpoint{1.535696in}{1.937123in}}%
\pgfusepath{stroke}%
\end{pgfscope}%
\begin{pgfscope}%
\pgfpathrectangle{\pgfqpoint{0.536528in}{0.633889in}}{\pgfqpoint{2.835476in}{1.604444in}}%
\pgfusepath{clip}%
\pgfsetbuttcap%
\pgfsetroundjoin%
\pgfsetlinewidth{1.003750pt}%
\definecolor{currentstroke}{rgb}{0.000000,0.000000,0.000000}%
\pgfsetstrokecolor{currentstroke}%
\pgfsetdash{}{0pt}%
\pgfpathmoveto{\pgfqpoint{1.704474in}{1.796559in}}%
\pgfusepath{stroke}%
\end{pgfscope}%
\begin{pgfscope}%
\pgfpathrectangle{\pgfqpoint{0.536528in}{0.633889in}}{\pgfqpoint{2.835476in}{1.604444in}}%
\pgfusepath{clip}%
\pgfsetbuttcap%
\pgfsetroundjoin%
\pgfsetlinewidth{1.003750pt}%
\definecolor{currentstroke}{rgb}{0.000000,0.000000,0.000000}%
\pgfsetstrokecolor{currentstroke}%
\pgfsetdash{}{0pt}%
\pgfpathmoveto{\pgfqpoint{2.210809in}{1.997365in}}%
\pgfusepath{stroke}%
\end{pgfscope}%
\begin{pgfscope}%
\pgfpathrectangle{\pgfqpoint{0.536528in}{0.633889in}}{\pgfqpoint{2.835476in}{1.604444in}}%
\pgfusepath{clip}%
\pgfsetbuttcap%
\pgfsetroundjoin%
\pgfsetlinewidth{1.003750pt}%
\definecolor{currentstroke}{rgb}{0.000000,0.000000,0.000000}%
\pgfsetstrokecolor{currentstroke}%
\pgfsetdash{}{0pt}%
\pgfpathmoveto{\pgfqpoint{2.379587in}{1.937123in}}%
\pgfusepath{stroke}%
\end{pgfscope}%
\begin{pgfscope}%
\pgfpathrectangle{\pgfqpoint{0.536528in}{0.633889in}}{\pgfqpoint{2.835476in}{1.604444in}}%
\pgfusepath{clip}%
\pgfsetbuttcap%
\pgfsetroundjoin%
\pgfsetlinewidth{1.003750pt}%
\definecolor{currentstroke}{rgb}{0.000000,0.000000,0.000000}%
\pgfsetstrokecolor{currentstroke}%
\pgfsetdash{}{0pt}%
\pgfpathmoveto{\pgfqpoint{2.885922in}{1.997365in}}%
\pgfusepath{stroke}%
\end{pgfscope}%
\begin{pgfscope}%
\pgfpathrectangle{\pgfqpoint{0.536528in}{0.633889in}}{\pgfqpoint{2.835476in}{1.604444in}}%
\pgfusepath{clip}%
\pgfsetbuttcap%
\pgfsetroundjoin%
\pgfsetlinewidth{1.003750pt}%
\definecolor{currentstroke}{rgb}{0.000000,0.000000,0.000000}%
\pgfsetstrokecolor{currentstroke}%
\pgfsetdash{}{0pt}%
\pgfpathmoveto{\pgfqpoint{3.054701in}{1.756398in}}%
\pgfusepath{stroke}%
\end{pgfscope}%
\begin{pgfscope}%
\pgfsetrectcap%
\pgfsetmiterjoin%
\pgfsetlinewidth{1.003750pt}%
\definecolor{currentstroke}{rgb}{0.000000,0.000000,0.000000}%
\pgfsetstrokecolor{currentstroke}%
\pgfsetdash{}{0pt}%
\pgfpathmoveto{\pgfqpoint{0.536528in}{0.633889in}}%
\pgfpathlineto{\pgfqpoint{0.536528in}{2.238333in}}%
\pgfusepath{stroke}%
\end{pgfscope}%
\begin{pgfscope}%
\pgfsetrectcap%
\pgfsetmiterjoin%
\pgfsetlinewidth{1.003750pt}%
\definecolor{currentstroke}{rgb}{0.000000,0.000000,0.000000}%
\pgfsetstrokecolor{currentstroke}%
\pgfsetdash{}{0pt}%
\pgfpathmoveto{\pgfqpoint{0.536528in}{0.633889in}}%
\pgfpathlineto{\pgfqpoint{3.372004in}{0.633889in}}%
\pgfusepath{stroke}%
\end{pgfscope}%
\begin{pgfscope}%
\definecolor{textcolor}{rgb}{0.000000,0.000000,0.000000}%
\pgfsetstrokecolor{textcolor}%
\pgfsetfillcolor{textcolor}%
\pgftext[x=0.674926in,y=1.766438in,,base]{\color{textcolor}\sffamily\fontsize{6.000000}{7.200000}\selectfont 58\%}%
\end{pgfscope}%
\begin{pgfscope}%
\definecolor{textcolor}{rgb}{0.400000,0.400000,0.400000}%
\pgfsetstrokecolor{textcolor}%
\pgfsetfillcolor{textcolor}%
\pgftext[x=1.215017in,y=1.505389in,,base]{\color{textcolor}\sffamily\fontsize{6.000000}{7.200000}\selectfont 45\%}%
\end{pgfscope}%
\begin{pgfscope}%
\definecolor{textcolor}{rgb}{0.000000,0.000000,0.000000}%
\pgfsetstrokecolor{textcolor}%
\pgfsetfillcolor{textcolor}%
\pgftext[x=1.350039in,y=1.907003in,,base]{\color{textcolor}\sffamily\fontsize{6.000000}{7.200000}\selectfont 65\%}%
\end{pgfscope}%
\begin{pgfscope}%
\definecolor{textcolor}{rgb}{0.400000,0.400000,0.400000}%
\pgfsetstrokecolor{textcolor}%
\pgfsetfillcolor{textcolor}%
\pgftext[x=1.890130in,y=1.766438in,,base]{\color{textcolor}\sffamily\fontsize{6.000000}{7.200000}\selectfont 58\%}%
\end{pgfscope}%
\begin{pgfscope}%
\definecolor{textcolor}{rgb}{0.000000,0.000000,0.000000}%
\pgfsetstrokecolor{textcolor}%
\pgfsetfillcolor{textcolor}%
\pgftext[x=2.025153in,y=1.967244in,,base]{\color{textcolor}\sffamily\fontsize{6.000000}{7.200000}\selectfont 68\%}%
\end{pgfscope}%
\begin{pgfscope}%
\definecolor{textcolor}{rgb}{0.400000,0.400000,0.400000}%
\pgfsetstrokecolor{textcolor}%
\pgfsetfillcolor{textcolor}%
\pgftext[x=2.565243in,y=1.907003in,,base]{\color{textcolor}\sffamily\fontsize{6.000000}{7.200000}\selectfont 65\%}%
\end{pgfscope}%
\begin{pgfscope}%
\definecolor{textcolor}{rgb}{0.000000,0.000000,0.000000}%
\pgfsetstrokecolor{textcolor}%
\pgfsetfillcolor{textcolor}%
\pgftext[x=2.700266in,y=1.967244in,,base]{\color{textcolor}\sffamily\fontsize{6.000000}{7.200000}\selectfont 68\%}%
\end{pgfscope}%
\begin{pgfscope}%
\definecolor{textcolor}{rgb}{0.400000,0.400000,0.400000}%
\pgfsetstrokecolor{textcolor}%
\pgfsetfillcolor{textcolor}%
\pgftext[x=3.240357in,y=1.726277in,,base]{\color{textcolor}\sffamily\fontsize{6.000000}{7.200000}\selectfont 56\%}%
\end{pgfscope}%
\begin{pgfscope}%
\definecolor{textcolor}{rgb}{0.000000,0.000000,0.000000}%
\pgfsetstrokecolor{textcolor}%
\pgfsetfillcolor{textcolor}%
\pgftext[x=1.954266in,y=2.321667in,,base]{\color{textcolor}\sffamily\fontsize{7.000000}{8.400000}\selectfont Model Evaluation on dataset of Miller R. et al.}%
\end{pgfscope}%
\begin{pgfscope}%
\pgfsetbuttcap%
\pgfsetmiterjoin%
\definecolor{currentfill}{rgb}{1.000000,1.000000,1.000000}%
\pgfsetfillcolor{currentfill}%
\pgfsetlinewidth{1.003750pt}%
\definecolor{currentstroke}{rgb}{1.000000,1.000000,1.000000}%
\pgfsetstrokecolor{currentstroke}%
\pgfsetdash{}{0pt}%
\pgfpathmoveto{\pgfqpoint{0.825670in}{0.201167in}}%
\pgfpathlineto{\pgfqpoint{3.082862in}{0.201167in}}%
\pgfpathlineto{\pgfqpoint{3.082862in}{0.365904in}}%
\pgfpathlineto{\pgfqpoint{0.825670in}{0.365904in}}%
\pgfpathlineto{\pgfqpoint{0.825670in}{0.201167in}}%
\pgfpathclose%
\pgfusepath{stroke,fill}%
\end{pgfscope}%
\begin{pgfscope}%
\pgfsetbuttcap%
\pgfsetmiterjoin%
\definecolor{currentfill}{rgb}{1.000000,0.650980,0.000000}%
\pgfsetfillcolor{currentfill}%
\pgfsetlinewidth{0.000000pt}%
\definecolor{currentstroke}{rgb}{0.000000,0.000000,0.000000}%
\pgfsetstrokecolor{currentstroke}%
\pgfsetstrokeopacity{0.000000}%
\pgfsetdash{}{0pt}%
\pgfpathmoveto{\pgfqpoint{0.864558in}{0.258960in}}%
\pgfpathlineto{\pgfqpoint{1.059003in}{0.258960in}}%
\pgfpathlineto{\pgfqpoint{1.059003in}{0.327015in}}%
\pgfpathlineto{\pgfqpoint{0.864558in}{0.327015in}}%
\pgfpathlineto{\pgfqpoint{0.864558in}{0.258960in}}%
\pgfpathclose%
\pgfusepath{fill}%
\end{pgfscope}%
\begin{pgfscope}%
\definecolor{textcolor}{rgb}{0.000000,0.000000,0.000000}%
\pgfsetstrokecolor{textcolor}%
\pgfsetfillcolor{textcolor}%
\pgftext[x=1.136781in,y=0.258960in,left,base]{\color{textcolor}\sffamily\fontsize{7.000000}{8.400000}\selectfont similarity-learning}%
\end{pgfscope}%
\begin{pgfscope}%
\pgfsetbuttcap%
\pgfsetmiterjoin%
\definecolor{currentfill}{rgb}{0.666667,0.666667,0.666667}%
\pgfsetfillcolor{currentfill}%
\pgfsetlinewidth{0.000000pt}%
\definecolor{currentstroke}{rgb}{0.000000,0.000000,0.000000}%
\pgfsetstrokecolor{currentstroke}%
\pgfsetstrokeopacity{0.000000}%
\pgfsetdash{}{0pt}%
\pgfpathmoveto{\pgfqpoint{1.953657in}{0.258960in}}%
\pgfpathlineto{\pgfqpoint{2.148102in}{0.258960in}}%
\pgfpathlineto{\pgfqpoint{2.148102in}{0.327015in}}%
\pgfpathlineto{\pgfqpoint{1.953657in}{0.327015in}}%
\pgfpathlineto{\pgfqpoint{1.953657in}{0.258960in}}%
\pgfpathclose%
\pgfusepath{fill}%
\end{pgfscope}%
\begin{pgfscope}%
\definecolor{textcolor}{rgb}{0.000000,0.000000,0.000000}%
\pgfsetstrokecolor{textcolor}%
\pgfsetfillcolor{textcolor}%
\pgftext[x=2.225879in,y=0.258960in,left,base]{\color{textcolor}\sffamily\fontsize{7.000000}{8.400000}\selectfont classification-learning}%
\end{pgfscope}%
\end{pgfpicture}%
\makeatother%
\endgroup%

%% file: sections/05_discussion.tex
\section{Discussion}
The question that must be asked first is whether the similarity-learning approach offers greater versatility compared to the prevalent classification-learning approach.
Both models demonstrated the capacity to generalize across sessions, tasks, and devices.
However, our findings reveal a notable difference in accuracy: the similarity-learning model surpassed the classification-learning baseline in accuracy, especially in contexts with limited enrollment data.
This characteristic is crucial for use cases where gathering comprehensive user data initially is not feasible. 
For instance, when collecting reference motions during initial user registration on a Social VR platform, it is unrealistic to expect users to provide such data over prolonged periods, e.g., 20 minutes.
Moreover, the capability to be pretrained and thus readily adapt to new users without comprehensive retraining is exclusive to the similarity-learning approach.
In contrast, the classification-learning approach lacks this key capability, considerably diminishing its viability for practical XR applications.
Therefore, similarity-learning models are very promising for developing practical motion-based user identification systems.
While not yet perfect, our findings make a compelling case for further research and refinement toward achieving fully versatile models that work across different use cases.

Crucially, the similarity-learning approach enables the practical application of motion-based user recognition in XR settings.
A pretrained similarity-learning model can be made available as a plugin for popular 3D engines like Unity or Unreal, mirroring the distribution model of current LLMs. 
For example, models such as LLAMA2 \cite{Touvron2023} or GPT \cite{Radford2018}, after being pretrained on vast datasets, are accessible for download and use across a variety of devices, irrespective of the user's technical expertise. 
This streamlined approach in deploying the similarity-learning model represents a substantial advancement in making motion-based recognition accessible and practical to XR researchers and practitioners. 

Furthermore, the similarity-learning model opens up new use cases, such as blacklists on social platforms:
should a user be banned for misconduct, their biometric template can be added to a blacklist, and the system ensures that any new user's identity is \textit{not} on this list.
This is difficult, if not impossible, to do with current approaches: feature-distance systems require a specific motion, which could be faked, and current classification-learning systems are not able to indicate that a user is unknown, so they fail as well.
The similarity-learning model, however, presents a viable solution:
if the embedding of a newcomer significantly diverges from all reference embeddings in the blacklist, it can be ascertained that the individual is not listed.


\section{Future Work}

Future research should concentrate on three key objectives to advance the field of motion-based recognition systems.
First, although our model exhibits greater versatility than traditional classification-learning models, further investigation is necessary to fully understand the scope of this versatility and identify opportunities for enhancement.
On a technical level, future research should explore improved pre-processing techniques to better isolate identifying signals and more sophisticated machine learning configurations and architectures.
Since our objective was not to fine-tune the models to achieve the highest possible performance on the dataset but rather to conduct a comparative and fair analysis between classification- and similarity-learning, we did not exhaust all avenues for optimization with our hyperparameter search.
More critically though, the field urgently requires larger and more diverse datasets not only for training but also for robust evaluation of these models.
Accordingly, future efforts should aim to compile datasets encompassing a broader array of users performing an even wider set of tasks than the dataset we have used.
We think it is also important to expand the demographic diversity of these datasets.
Most publicly available motion datasets currently feature relatively young users predominantly from the US or Germany.
To achieve genuine versatility and global applicability, models must be trained and validated on data that reflects a better representation of the world population.

Second, efforts should be directed towards transitioning these models from research to production, building on the foundational work established with our similarity-learning approach.
For instance, our group has started developing a Unity Plugin and a corresponding server backend, which we plan to evaluate and detail in forthcoming publications.
This client-server architecture enables the XR application to transmit motion sequences to the server, where a pretrained similarity-learning model processes these sequences.
In turn, the server manages the resulting user embeddings and returns the outcome of identification and verification requests back to the XR application.
This design ensures minimal requirements for both the XR developer's machine learning skills and the system's hardware, as the model is already trained.
However, given the sensitive nature of both the motion data and the resulting embeddings, future developments must implement adequate security measures to protect the privacy and safety of user data in these systems.

Lastly, while this work is largely motivated by the positive aspects of what motion data can be used for, we also note that motion data is more than just a fingerprint \cite{NairTruthMotionUnprecedented2023}.
Due to its dynamic properties, motion data not only conveys obvious information about body proportions \cite{NairTruthMotionUnprecedented2023} but also about users' internal states and emotions~\cite{venture2014recognizing} and about sensitive attributes, such as gender, age, or fitness \cite{NairInferringPrivatePersonal2024}.
Overall, movement data appears to be a highly sensitive asset also worth protecting.
Notably, this sensitive data is shared across other users' remote XR devices in multi-user environments.
This is a privacy issue worth investigating, as there currently is a significant lack of individual control over one's sensitive data or even safeguarding mechanisms against avatar-based fraud attempts in these architectures.
Current research has already started exploring potential countermeasures \cite{NairGoingIncognitoMetaverse2023, NairDeepMotionMasking2023}, and we encourage future work to not only focus on novel motion-based identification and verification solutions but also continue pursuing these important privacy questions.

\section{Conclusion}
Our work has highlighted the potential of pretrained similarity-learning models to facilitate user identification and verification in XR environments.
By separating users into training, validation, and testing sets, we demonstrated that the similarity-learning model can be pretrained on one set of users and then immediately applied for the identification task of unknown users.
This capability represents a noteworthy advancement over previous studies \cite{Miller2021}, which did not demonstrate this crucial aspect of model versatility.
These findings lay the groundwork for the practical deployment of motion-based user recognition in production XR systems.
While this represents an important initial step, the versatility of these models must be further enhanced and examined.
Future research should ensure these models are effective across a diverse range of user tasks and adaptable to users from various global origins.
Looking ahead, we expect ongoing research and development efforts to continue refining these approaches, thereby establishing motion-based biometric signals as an important element of security in the immersive experiences of the future.